\documentclass[sigconf]{acmart}

\AtBeginDocument{%
  \providecommand\BibTeX{{%
    \normalfont B\kern-0.5em{\scshape i\kern-0.25em b}\kern-0.8em\TeX}}}

\copyrightyear{2026}
\acmYear{2026}
\setcopyright{cc}
\setcctype{by-nc-nd}
\acmConference[CHI '26]{Proceedings of the 2026 CHI Conference on Human Factors in Computing Systems}{April 13--17, 2026}{Barcelona, Spain}
\acmBooktitle{Proceedings of the 2026 CHI Conference on Human Factors in Computing Systems (CHI '26), April 13--17, 2026, Barcelona, Spain}
\acmPrice{}
\acmDOI{10.1145/3772318.3790341}
\acmISBN{979-8-4007-2278-3/2026/04}

\usepackage{makecell}
\usepackage{multirow}
\usepackage{mathtools}
\usepackage{listings}
\usepackage{verbatim}
\usepackage[utf8]{inputenc}
\usepackage{enumitem}
\usepackage{colortbl}
\usepackage{array}

\usepackage[linesnumbered,ruled,vlined]{algorithm2e}
\begin{document}

\title{Behavior-Aware Anthropometric Scene Generation for Human-Usable 3D Layouts}

\author{Semin Jin}
\authornote{Co-first authors.}
\affiliation{%
  \department{Design Informatics Lab}
  \institution{Hanyang University}
  \city{Seoul}
  \country{Republic of Korea}
}
\affiliation{%
  \department{Human-Centered AI Design Institute}
  \institution{Hanyang University}
  \city{Seoul}
  \country{Republic of Korea}
}
\email{tpals97@gmail.com}

\author{Donghyuk Kim}
\authornotemark[1]
\affiliation{%
  \department{Design Informatics Lab}
  \institution{Hanyang University}
  \city{Seoul}
  \country{Republic of Korea}
}
\affiliation{%
  \department{Human-Centered AI Design Institute}
  \institution{Hanyang University}
  \city{Seoul}
  \country{Republic of Korea}
}
\email{oververitas@gmail.com}

\author{Jeongmin Ryu}
\affiliation{%
  \department{Design Informatics Lab}
  \institution{Hanyang University}
  \city{Seoul}
  \country{Republic of Korea}
}
\email{2002rjm@gmail.com}

\author{Kyung Hoon Hyun}
\authornote{Corresponding author.}
\affiliation{%
  \department{Design Informatics Lab}
  \institution{Hanyang University}
  \city{Seoul}
  \country{Republic of Korea}
}
\affiliation{%
  \department{Human-Centered AI Design Institute}
  \institution{Hanyang University}
  \city{Seoul}
  \country{Republic of Korea}
}
\email{hoonhello@gmail.com}

\begin{abstract}
Well-designed indoor scenes should prioritize how people can act within a space rather than merely what objects to place. However, existing 3D scene generation methods emphasize visual and semantic plausibility, while insufficiently addressing whether people can comfortably walk, sit, or manipulate objects. To bridge this gap, we present a Behavior-Aware Anthropometric Scene Generation framework. Our approach leverages vision–language models (VLMs) to analyze object–behavior relationships, translating spatial requirements into parametric layout constraints adapted to user-specific anthropometric data. We conducted comparative studies with state-of-the-art models using geometric metrics and a user perception study (\textbf{N}=16). We further conducted in-depth human-scale studies (individuals, \textbf{N}=20; groups, \textbf{N}=18). The results showed improvements in task completion time, trajectory efficiency, and human-object manipulation space. This study contributes a framework that bridges VLM-based interaction reasoning with anthropometric constraints, validated through both technical metrics and real-scale human usability studies.
\end{abstract}




\begin{teaserfigure}
\centering
\includegraphics[width=\textwidth]{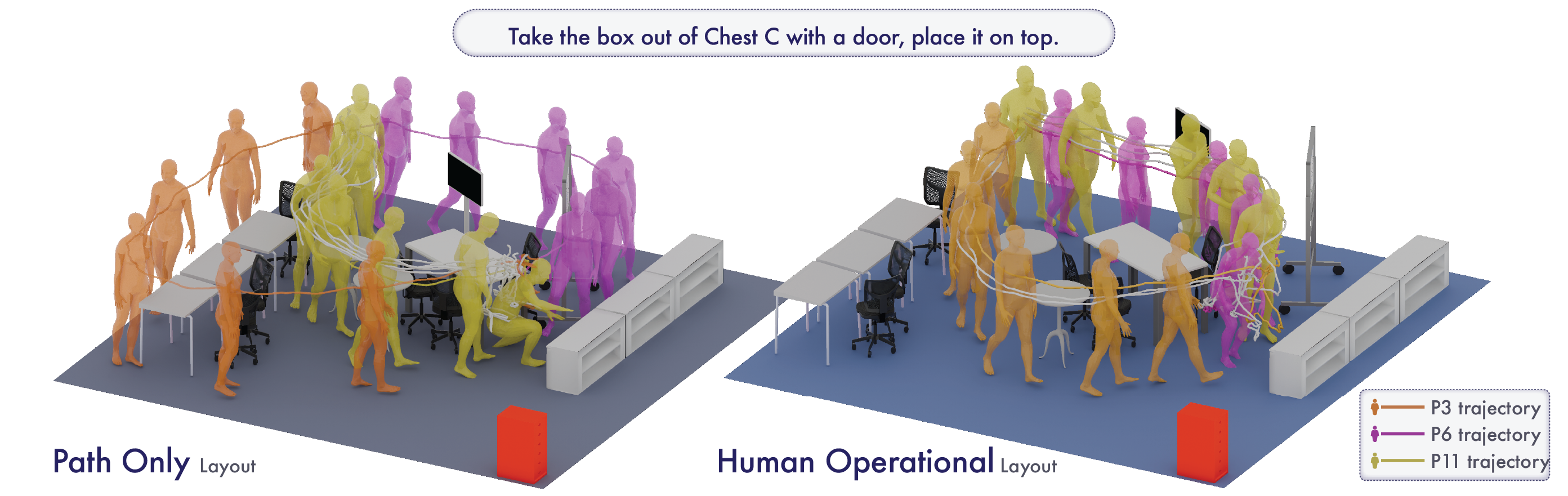}
\caption{Visualization of Movement Trajectories in Generated Path-Only and Human-Operational Layouts.}
\Description{A visualization compares movement paths between two conditions. The left panel, labeled Human Operational Layout, shows participants moving in straight, efficient lines around furniture. In contrast, the right panel, labeled Path-Only Layout, shows the same participants taking inefficient and circuitous paths with noticeable detours, visually highlighting the navigation inefficiency caused by the layout.}
\label{fig:1}
\end{teaserfigure}

\maketitle

\section{Introduction}
Physical environments fundamentally shape human movements and behavior \cite{luo2023pearl}. In the real world, layouts are rarely static; users naturally adjust their surroundings—shifting a chair back to create legroom or pulling a table closer to reach an object—to fit their specific body dimensions and movements. A usable layout is then one already optimized for its occupants' anthropometrics and behaviors. Recent progress in 3D scene generation has enabled automatic creation of plausible layouts \cite{ccelen2024design, lin2024instructscene, paschalidou2021atiss, sun2025layoutvlm, wang2021sceneformer}. However, a critical limitation is that these models generate layouts based on dataset statistics \cite{lin2024instructscene, paschalidou2021atiss, wang2021sceneformer} or large language model (LLM) priors \cite{ccelen2024design, sun2025layoutvlm}, lacking any explicit reasoning about human dimensions or behavioral patterns. 

Consider an office scenario: a generative model might place desks back-to-back based on visual symmetry found in training dataset. However, because the model ignores the anthropometric clearance required to actually push a chair back and stand up, the resulting layout creates an immediate conflict zone—a functional failure that a human user would have instinctively avoided by adjusting the furniture distance. This problem becomes even more critical in extended reality (XR) applications requiring embodied interaction fidelity, such as training simulations, telepresence, or digital twins for ergonomic assessment. For example, users may repeatedly reposition themselves to reach targets outside their movement range, or experience visual clipping where virtual hands penetrate surfaces—issues arising because the layout was not optimized for user-specific body dimensions. Although established design standards address functional constraints such as drawer clearances and circulation widths, applying them to layout generation is challenging: the guidelines specify recommended ranges that cannot be reduced to fixed values without specific contexts, and generating a usable layout requires holistic consideration of object types, dimensions, room configurations, and user body characteristics. Our framework addresses these challenges by leveraging VLM reasoning to infer object-specific spatial constraints and grounding them in individualized anthropometric data.

To address this limitation, we propose \textbf{Behavior-Aware Anthropometric Scene Generation}, a vision language model (VLM)-based framework that augments text-based layout generation using \textit{behavioral reasoning} and \textit{anthropometric grounding}. We leverage VLMs to infer object functions from visual cues and reason about potential human interactions based on scene type and layout criteria. Our framework uses layout criteria and assets as inputs and proceeds in two stages. First, it constructs behavior-aware relational representations that integrate object semantics, human–object interaction patterns, and group-level spatial relations. Second, it performs constraint-based layout generation by inferring anthropometrically grounded spatial constraints from these relations and encoding them as differentiable penalty terms for gradient-based optimization. By parameterizing these constraints with personalized anthropometric profiles, our approach ensures that the resulting scenes support functional connectivity, adequate operational clearances, and efficient circulation paths tailored to individual users.

We evaluated our framework through technical validation and user studies on human-operational usability. First, we conducted performance comparisons with state-of-the-art LLM-based scene generation methods, LayoutVLM \cite{sun2025layoutvlm}, to measure object collisions, floor-plan violations, and professionals' perceptions of generated layouts (N = 16). Second, we implemented the generated layouts at a 1:1 scale in physical office and lounge environments and compared three conditions: \textbf{LayoutVLM (Baseline),} \textbf{Passage-Only (PO)}, which ensures minimal navigable passages using static body dimensions; and  \textbf{Human-Operational (HO)}, which guarantees sufficient human-operational space based on movement-related anthropometric envelopes. For PO and HO conditions, we instantiated participant-specific anthropometric profiles from each participant’s Skinned Multi-Person Linear (SMPL) \cite{SMPL:2015} model, parameterizing the constraints (e.g., passage widths, reaching envelopes, and viewing requirements) to each participant’s actual body dimensions. Our user studies on human-operational usability consist of two parts: an individual study (N = 20) using structured \textit{object--action--target} tasks \cite{achlioptas2020referit3d, wang2022humanise} to validate the anthropometrically grounded layout support task performance, and a group study (N = 18, six teams of three) using naturalistic collaborative scenarios to examine whether layouts minimize circulation conflicts during shared tasks \cite{luo2025documents, luo2022should}.

The key contributions of our research include:
\begin{itemize}
    \item We propose a behavior-aware anthropometric scene generation framework that leverages VLM reasoning to infer spatial constraints from human-object interactions and grounds them in individual anthropometric data.
    \item We instantiated these constraints within a differentiable layout optimization process, operationalizing passage widths, operational clearances, and interaction-space occupancy to generate human-operational layouts.
    \item We conducted an evaluation combining technical validation, professional perception study, and user studies on human-operational usability in real-scale environments.
\end{itemize}

\section{Related Works}
Recent 3D scene generation methods synthesize plausible layouts but rely on generic spatial constraints that overlook individual body dimensions and behavioral patterns. Generating human-usable layouts requires anthropometric design principles, yet existing evaluations rarely assess whether generated layouts support actual human operation. To address these gaps, we review prior work in three areas: LLM-based scene generation, anthropometric design in virtual and physical spaces, and evaluation methodologies for 3D layouts.

\subsection{LLM-based Scene Generation}
Recent advances in 3D indoor scene generation have leveraged Transformer architectures \cite{wang2021sceneformer, paschalidou2021atiss, sun2025forest2seq}, diffusion models \cite{lin2024instructscene, zhai2023commonscenes, Tang_2024_CVPR}, and LLMs \cite{sun2025layoutvlm, ccelen2024design, yang2024holodecklanguageguidedgeneration} to generate semantically plausible layouts. Among these approaches, LayoutVLM \cite{sun2025layoutvlm} represents the current state-of-the-art in LLM-based scene generation by formulating layout creation as a constraint-satisfaction problem. It leverages LLMs to translate natural-language instructions into spatial constraints (e.g., distance, orientation, and alignment) and then optimizes object positions through gradient-based methods using visual language model feedback. This approach demonstrates strong performance in generating semantically coherent layouts that satisfy user-specified relationships. However, LayoutVLM's reliance on LLM common sense for distance parameters results in generic, one-size-fits-all solutions that fail to consider individual body dimensions or behavioral requirements. 

A separate line of research explores LLMs for human behavior simulation within existing scenarios \cite{qu2024gpt, su2023scene, wu2025human, ding2023task}. These approaches excel at generating human motions and planning action sequences in pre-defined spatial layouts but differ from our objective: they adapt human behavior to fit existing spaces, whereas we generate spaces to fit human-operational needs. Our work builds upon LayoutVLM's constraint optimization framework but extends it by integrating anthropometric data directly into the constraint quantification process. Rather than relying on generic distances from LLM common sense, we compute person-specific operational requirements based on individual body measurements and intended interactions--advancing scene generation from semantically plausible to human-operational.

\subsection{Anthropometric-Driven Design in Virtual and Physical Scenes}
Foundational theories in architecture and human factors establish that spatial design is not merely visual but fundamentally interaction-oriented. In architectural theory, the relationship between the human body and space is central, where operational space is defined by the dynamic range required for human activity \cite{damon1966human, panero1979human, ching2023architecture}. Similarly, ergonomics literature distinguishes between structural anthropometry---static body dimensions---and functional anthropometry, which describes the dynamic range of motion and clearance required for tasks, emphasizing that true usability depends on accommodating the latter \cite{viviani2018accuracy}. Aligning with these theoretical frameworks, research in ergonomics and XR demonstrates the critical importance of anthropometric considerations in the functional space. Research on personalized furniture design has shown that incorporating individual body measurements—such as reach envelope and joint range of motion—directly improves task performance \cite{ramakers2023measurement}. XR environments have further embraced this approach, in which virtual objects dynamically adapt to user-body dimensions for optimal interaction \cite{lee2018interactive, lee2016posing, chen2025exploring}. These systems adjust shelf heights based on arm reach, scale workspaces to accommodate sitting eye height, and position controls within comfortable manipulation zones. However, these approaches primarily evaluate the design process itself, measuring whether the generated furniture dimensions match body measurements, rather than assessing actual human behavior in the resulting spaces. Consequently, the gap between anthropometric specifications and real-world use remains largely unknown, and the cascading effects on human-scene and human-object interactions have not been systematically studied in the context of scene generation.

The anthropometric design approach extends beyond individual furniture pieces to encompass operational spaces and the volume required for humans to use objects effectively. The established guidelines \cite{panero1979human, dianat2018review} define the clearances for drawer operations, circulation paths around desks, and viewing distances. However, current scene generation methods treat these as fixed constants. Our human-operational approach bridges this gap by translating anthropometric measurements into spatial constraints that ensure adequate operational space for each individual's body dimensions. This shift from generic clearances to personalized operational volumes represents an advancement in making generated scenes truly usable, rather than merely plausible. The importance of this personalization becomes evident when considering human diversity; a 5th percentile female and a 95th percentile male require substantially different operational spaces \cite{panero1979human}; however, current methods apply uniform constraints that may be insufficient for larger individuals or wastefully spacious for smaller ones. By integrating anthropometric data directly into the generation process, we ensure that layouts accommodate specific individuals who will inhabit these spaces.

\subsection{3D Scene Evaluation Methods}
Existing 3D scene generation research has primarily relied on visual plausibility metrics such as Fréchet Inception Distance and Kernel Inception Distance \cite{paschalidou2021atiss, Tang_2024_CVPR, lin2024instructscene} to measure the distributional similarity to training data, or user ratings \cite{ccelen2024design, sun2025layoutvlm, su2023scene} to assess semantic coherence. Although these metrics effectively validate visual quality and learning performance, they fail to capture the behavioral and operational aspects that determine the actual usability. In contrast, architecture and interior design fields employ a comprehensive post-occupation evaluation that assesses cognitive comfort, way-finding efficiency, spatial satisfaction, and long-term behavioral adaptation \cite{julia2021waiting, julia2020spatial, nguyen2024adaptive}; however, these require extended observation periods and are impractical for evaluating generated scenes at scale. Recent simulation-based approaches have attempted to bridge this gap by evaluating furniture mobility and walkability using embodied artificial intelligence (AI) agents \cite{yang2024physcene}. However, these methods cannot capture human behavioral flexibility, as people dynamically adapt by stepping over obstacles, crouching to pass through tight spaces, or creating unexpected optimal paths. Motion generation research \cite{wang2024move, wang2022humanise} models human-scene interactions more realistically, but requires extensive training data and struggles with behavioral improvisation.

\begin{figure*}[t]
\centering
\includegraphics[width=\textwidth]{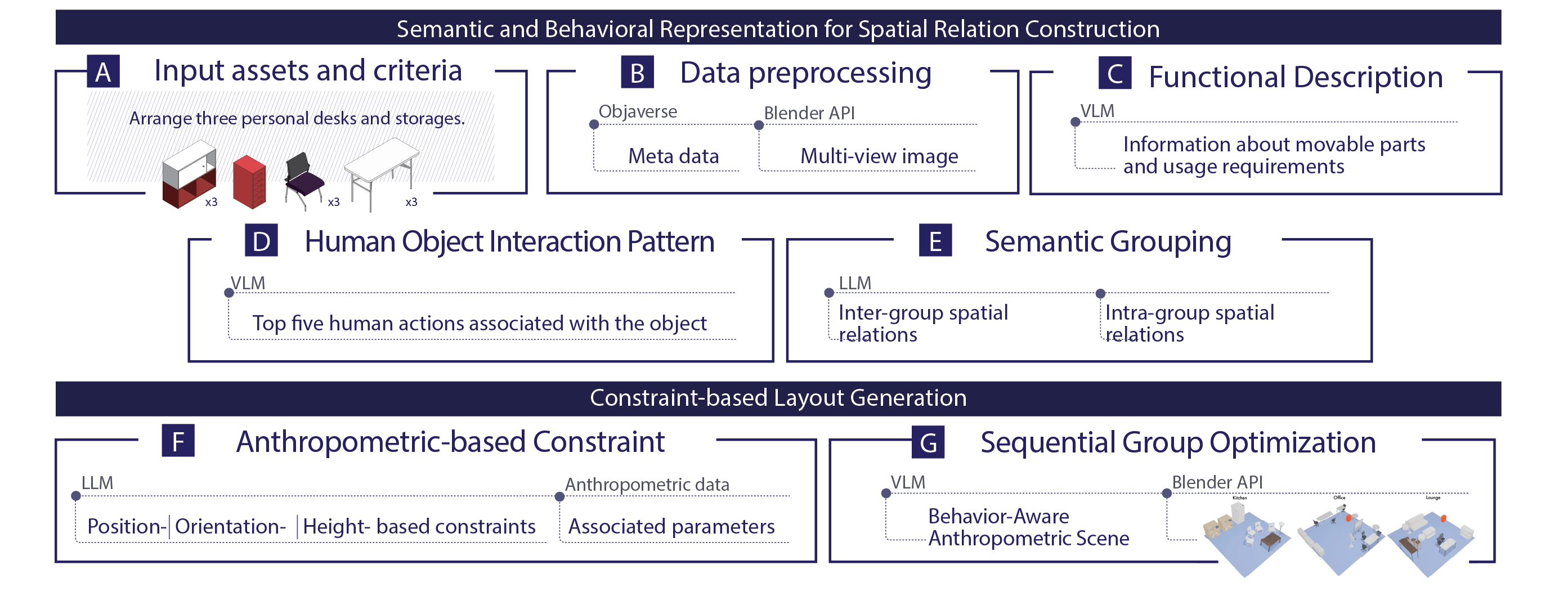}
\caption{Overview of the Behavior-Aware Anthropometric Scene Generation. The framework proceeds in two phases: Semantic and Behavioral Representation (Stages A–E) constructs spatial relations, and Anthropometric Constraint-based Layout Generation (Stages F–G) optimizes the final layout using anthropometric constraints.}
\Description{A diagram presents an overview of the Behavior-Aware Anthropometric Scene Generation framework. The framework proceeds in two phases: Semantic and Behavioral Representation (Steps A–E) constructs spatial relations, and Constraint-based Layout Generation (Steps F–G) optimizes the final layout using anthropometric constraints.
}
\label{fig:2}
\end{figure*}

To address this evaluation gap between visual metrics and real-world usability, we developed behavior-grounded metrics specifically designed for the generated 3D layouts. These metrics measure how humans navigate and utilize space. Our approach captures observable behavioral patterns: trajectory variability reveals layout intuitiveness, action sequences expose compensatory movements, and occupancy ratios validate operational space adequacy. This evaluation framework provides empirical evidence that anthropometric-aware generation improves functional performance, establishing a new paradigm for assessing generated scenes based on human-operational criteria.

\section{Behavior-Aware Anthropometric Scene Generation}

We present \textbf{Behavior-Aware Anthropometric Scene Generation}, an approach that augments language-based layout generation using behavioral reasoning and anthropometric grounding. The goal is to generate spatial layouts that align with both furniture function and human-object interaction. Our approach enables the system to infer spatial constraints by explicitly referencing behavioral context and anthropometric data. As shown in Figure~\ref{fig:2}, our framework uses scene instructions and assets as inputs and proceeds through two main stages:
\begin{enumerate}
\item \textbf{Semantic and Behavioral Representation for Spatial Relation Construction}: Constructing behavior-aware relational representations that integrate object semantics, human-object interaction patterns, and group-level spatial relations (Figure~\ref{fig:2}A--E).
\item \textbf{Constraint-based Layout Generation}: Inferring anthropometrically grounded constraint representations suitable for differentiable spatial optimization (Figure~\ref{fig:2}F--G).
\end{enumerate}
Complete prompt templates are provided in Appendix A for reproducibility.

\subsection{Semantic and Behavioral Representation for Spatial Relation Construction }
\label {sec:3.1}
The \textbf{[A-E] stages} interpret raw 3D assets and layout criteria to produce behavior-aware relational representations that link object geometry, function, and human interaction. These representations establish the contextual foundation required for constraint inference by modeling how objects are used, accessed, and coordinated within a scene. The following sections describe how we organize the asset data, extract behavioral features, and semantically group objects to prepare for constraint inference.

\begin{figure*}[t]
\centering
\includegraphics[width=\textwidth]{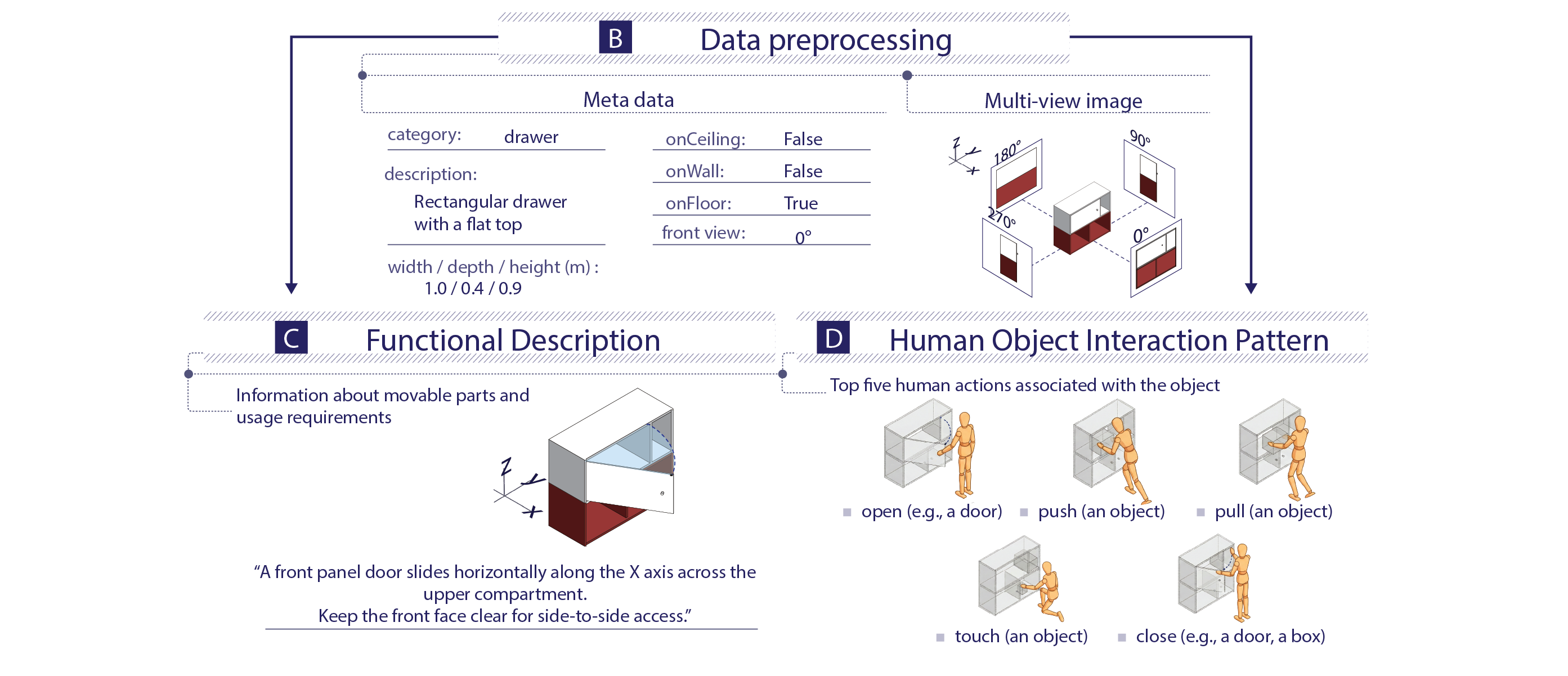}
\caption{Semantic and Behavioral Representation (Stages B–D). For every 3D asset used in the scene, [B] preprocesses metadata and multi-view renderings, [C] infers a functional description, and [D] extracts a human–object interaction pattern.}
\Description{A diagram shows the framework for semantic and behavioral representation. Stage [B] Data Preprocessing extracts metadata from 3D assets and generates multi-view images. Stage [C] Functional Description uses VLM to analyze movable parts and spatial requirements from these images. Stage [D] Human-Object Interaction Pattern identifies the top five human actions associated with each object based on visual analysis.}
\label{fig:3}
\end{figure*}

\subsubsection{Data and Input}
Given user-specified layout criteria (e.g., room type, required furniture), we retrieved 3D assets from open-universe datasets and placed them into initial scene layouts: \textbf{[A] Input assets and criteria }(Figure~\ref{fig:2}A). For each input asset, we leverage OpenShape \cite{liu2023openshape}, a large-scale multimodal model that retrieves 3D assets by aligning point clouds with text and image descriptions, to extract candidate objects from the open-universe dataset Objaverse \cite{deitke2023objaverse}. Once assets are determined, we extract each object's 3D bounding box, metadata (category, width, depth, height), and generate multi-view renderings of the 3D model using a scripted Blender pipeline: \textbf{[B] Data Preprocessing} (Figure~\ref{fig:3}B). Local coordinates were standardized (+X: right, +Y: forward, +Z: upward). We employed a VLM (GPT-4o) to interpret visual and textual cues jointly through prompts describing an object's geometry, parts, and possible interactions. To provide the VLM with visual grounding, we implemented an automated rendering pipeline using Blender. Each object was rendered from four orthogonal viewpoints (0$^{\circ}$, 90$^{\circ}$, 180$^{\circ}$, and 270$^{\circ}$) to capture comprehensive geometric details. The multi-view approach captures fine-grained features (e.g., casters, hinges, or door knobs) that dictate functionality but are often absent from text descriptions. We paired these multi-view images with structured metadata, enabling the VLM to cross-reference visual evidence with textual descriptions. The combined input enables the system to infer the functional properties of an object, such as movement axes, articulation points, and kinematic constraints, which are not evident in static category labels.

\subsubsection{Human-Object Interaction-based Feature Extraction}
Object geometry alone does not specify how furniture should be accessed or operated--a cabinet may open outward or slide laterally, and a chair may swivel or remain fixed. This stage infers each object's functional and behavioral properties: how it operates and how humans can interact with it. The VLM takes the paired multi-view image set and metadata as input and produces two complementary outputs: \textbf{[C] Functional Description} (Figure~\ref{fig:3}C), capturing qualitative information about movable parts and usage requirements, and \textbf{[D] Human-Object Interaction Pattern} (Figure~\ref{fig:3}D), identifying the top five human actions associated with the object based on atomic visual actions (e.g., sit, open, pull)~\cite{gu2018ava}. The resulting vision-to-text process produces a structured JSON representation that links the objects to their inferred functions and interaction semantics. These representations provide a behavioral foundation for constraint inference (detailed prompts in Figure A1; output example in Figure A2).

\subsubsection{Semantic Grouping}
\begin{figure*}[t]
\centering
\includegraphics[width=\textwidth]{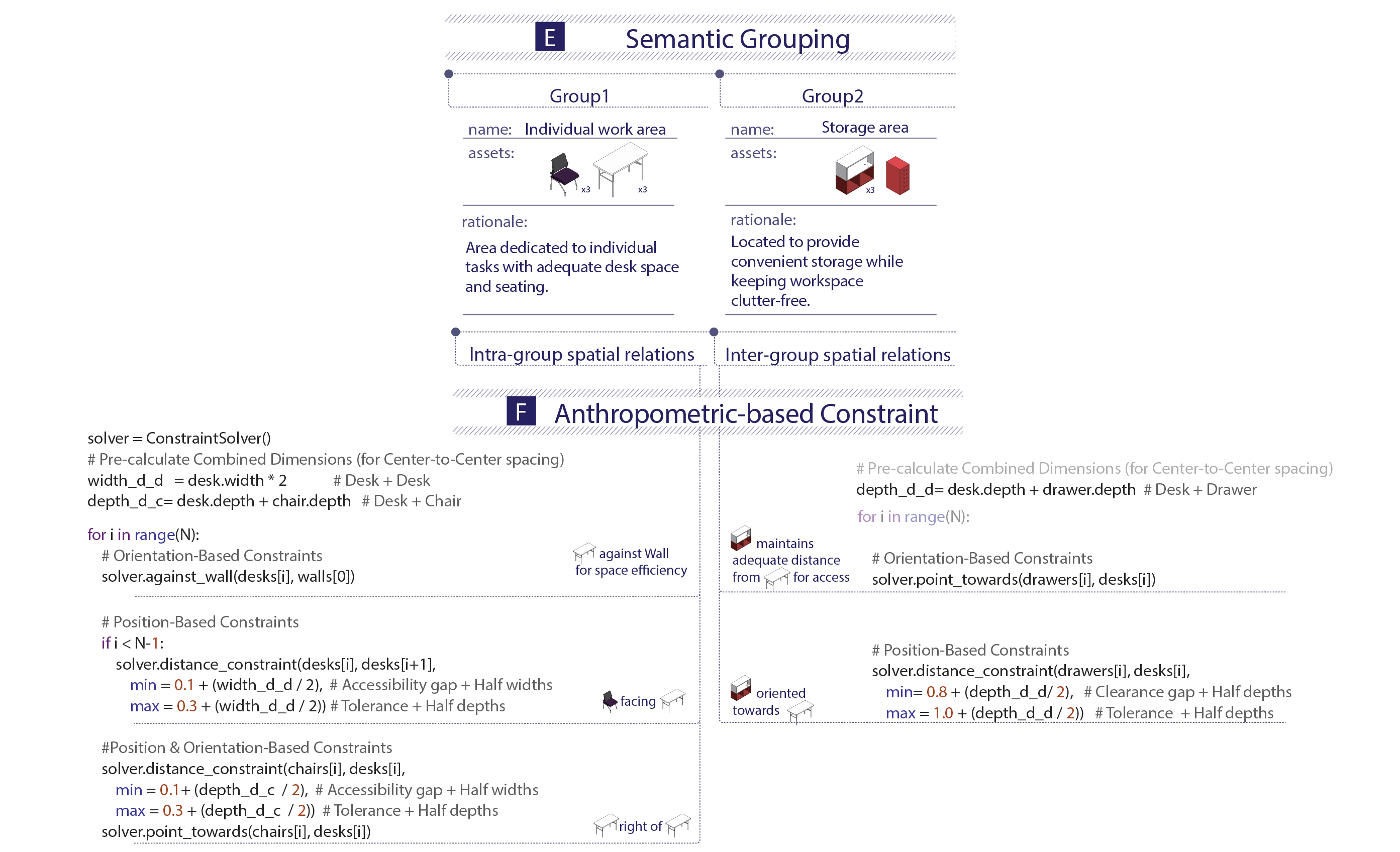}
\caption{[E] Semantic Grouping and [F] Anthropometric-based Constraint Inference.}
\Description{A diagram illustrates Stage [E] Semantic Grouping, which organizes assets into functional groups like individual work areas and storage areas. It defines intra-group relations for internal arrangement and inter-group relations to maintain accessibility between groups. Stage [F] Anthropometric-based Constraint applies body dimensions to calculate spatial requirements. Position-based constraints ensure accessibility gaps through half-width and half-depth calculations. Orientation-based constraints solve against wall distances and desk orientations to enable efficient interaction.}
\label{fig:4}
\end{figure*}

The \textbf{[E] Semantic Grouping} (Figure~\ref{fig:4}E) stage organizes objects into functional groups that reflect how humans interact with them in a scene. While individual objects can be interpreted in isolation, meaningful spatial reasoning emerges when they are considered as part of behavioral configurations; for example, chairs around a desk forming \textit{a workspace} or sofas arranged around a coffee table creating \textit{a lounge area.} Semantic grouping has two purposes. First, it reinterprets the atomic human-object interactions inferred in Stage [D] within a group context to extract spatial definitions. Specifically, the system identifies both \textbf{intra-group spatial relations} (internal arrangement within a functional unit) and \textbf{inter-group spatial relations} (connectivity between distinct groups), which are subsequently converted into a structured symbolic program in Stage [F]. For instance, an atomic `open' or `pull (an object)' action becomes the higher-level relation `organize' when understood within a multi-cabinet storage group. Second, it establishes a structural abstraction that reduces the optimization complexity in Stage [G] by treating each group as an independent unit with a behavioral priority determined by functional significance and object scale (e.g., scene-defining elements such as beds or desks are placed first). We performed semantic asset grouping using the system in Appendix Figure A3, defining each group as a set of objects linked by functional relation, geometric proximity, and shared human action (output example in Figure A4).

\subsection{Constraint-based Layout Generation}
\label {sec:3.2}
The \textbf{[F-G]} stages infer anthropometrically grounded constraint representations from natural-language spatial relations. They convert these relations, derived from the object geometry, function, and human-object interaction, into executable, differentiable constraints required for spatial optimization. Unlike prior layout approaches that rely on LLM common sense \cite{sun2025layoutvlm, feng2023layoutgpt, yang2024holodecklanguageguidedgeneration}, our framework infers constraint specifications by explicitly referencing behavioral semantics and anthropometric rationale.

\subsubsection{Spatial Constraint Definition}
To define the spatial relationships between objects, we constructed an extended taxonomy adapted from geometric relation formulations used in prior layouts and scene generation methods \cite{sun2025layoutvlm, yang2024holodecklanguageguidedgeneration, paschalidou2021atiss}. We then reinterpreted the constraint semantics and parameterization in a behaviorally and anthropometrically grounded manner. Prior work defines constraints in geometric terms, focusing on the distances and angles between assets, independent of how people interact with them. By contrast, our formulation explicitly encodes how people need to reach, access, and operate furniture. The taxonomy organizes natural-language relations into learnable constraint types---positional, orientational, and height-based---such as \textit{chair against wall, table aligned with sofa,} or \textit{lamp on top of desk.} Table~\ref{tab:taxonomy} lists each constraint type and the corresponding constraint names.

Among these, \textit{distance constraint} requires additional clarification, because it directly governs human accessibility and functional clearance. We represent each distance constraint as a center-to-center distance range [$d_{\min}$, $d_{\max}$], explicitly derived from the manner in which people reach, stand, and move around objects. Here, 
$d_{\text{accessibility}}$ represents the minimum distance required to reach or access an object, while $d_{\text{clearance}}$ represents the space needed for operational movements (e.g., pulling out a chair). We infer these bounds based on the interaction semantics: for accessibility-focused relations, $d_{\min} = d_{\text{accessibility}}$ and $d_{\max} = d_{\min} + \tau$; for clearance-focused relations, $d_{\min} = d_{\text{clearance}} - \tau$ and $d_{\max} = d_{\text{clearance}}$, where $\tau$ is a tolerance buffer inferred by the VLM based on object function and interaction context.

\subsubsection{Anthropometric-based Constraint Inference}
In the \textbf{[F] Anthropometric based Constraint Inference} stage (Figure~\ref{fig:4}F), we infer a complete constraint specification, including the constraint type, its parameters, and the associated anthropometric rationale, from natural-language spatial relations. We first parse each relation (e.g., \textit{Office Chair facing Desk} for intra-group relations, \textit{Double Chest maintains adequate distance from Desk} for inter-group relations) and assign it to one of the predefined constraint types. Each type includes a parameter template that specifies the spatial quantities to be inferred, such as distances, angles, or vertical offsets. To infer these parameters, we reference standardized anthropometric datasets using \textit{Human Dimension and Interior Space} \cite{panero1979human}, which aggregate data from multiple anthropometric sources \cite{damon1966human, nasa1978anthropometric, van1972human}. These datasets serve as population-level references. We use the 5th–95th percentile ranges for key horizontal dimensions, including forward reach, lateral reach, body breadth, and body depth, to represent the natural variability across the population. This enables our inferred constraints to remain valid across diverse body sizes and movement capabilities, without assuming a single average individual.

We then mapped each constraint type to a specific anthropometric rationale: reach-related relations use arm-reach measures, clearance and circulation relations use body breadth or depth, adjacency relations use minimal offsets required for co-functioning objects, and visual/orientation constraints use eye position and preferred viewing directions to align object fronts toward the primary interaction or viewing area. By integrating the relation type, the relevant anthropometric ranges, and the interactions, we infer the constraint type and its associated parameters such as \texttt{distance constraints} [$d_{min}, d_{max}$], \texttt{align\_0$^{\circ}$}, and \texttt{align\_90$^{\circ}$}. Finally, we encode this specification in an executable optimization schema that records the constraint type, object pair, parameter values, and anthropometric rationale. This structured representation is passed directly to the differentiable optimization module in Stage [G] (system prompt in Appendix Figure A5; output example in Figure A6).

\begin{table*}[t]
\centering
\caption{Spatial Constraint Taxonomy for Behavior-Aware Anthropometric Scene Generation, comparing our behavior-aware, anthropometrically grounded constraints with the geometric relations~\cite{sun2025layoutvlm}.}
\label{tab:taxonomy}
\small

\renewcommand{\arraystretch}{1.35}

\begin{tabular}{
    >{\centering\arraybackslash}m{2.2cm}
    >{\centering\arraybackslash}m{3.0cm}
    >{\centering\arraybackslash}m{2.3cm}
    p{9.0cm}
}
\toprule
\textbf{Constraint Type} &
\textbf{Constraint Name} &
\textbf{Method} &
\textbf{Description} \\
\midrule

\multirow{4}{*}[-1.2em]{Position-based} 

& $L_{\text{distance}}(p_i, p_j, d_{\min}, d_{\max})$
& \cellcolor{gray!10} LayoutVLM~\cite{sun2025layoutvlm}
& \cellcolor{gray!10} Distance between the two assets should fall within the range $[d_{\min}, d_{\max}]$. \\

&
& Ours
& Distance between two objects to $[d_{\min}, d_{\max}]$, where bounds are inferred from reach and clearance requirements based on anthropometric data. \\

& $L_{\text{against wall}}(p_i, w_j, b_i)$
& \cellcolor{gray!10} LayoutVLM~\cite{sun2025layoutvlm}
& \cellcolor{gray!10} Place an asset against wall $w_j$. \\

&
& Ours
& Places the object against a specific wall while considering accessibility and clearance requirements for nearby interactions. \\
\midrule

\multirow{4}{*}[-1.2em]{Orientation-based} 

& $L_{\text{align with}}(p_i, p_j, \Theta)$
& \cellcolor{gray!10} LayoutVLM~\cite{sun2025layoutvlm}
& \cellcolor{gray!10} Align two assets at a specified angle $\Theta$. \\

&
& Ours
& Aligns the rotations of two objects; the angle parameter $\Theta$ reflects task-oriented alignment (e.g., parallel or perpendicular configurations for joint use). \\

& $L_{\text{point towards}}(p_i, p_j, \Theta)$
& \cellcolor{gray!10} LayoutVLM~\cite{sun2025layoutvlm}
& \cellcolor{gray!10} Orient one asset to face another with an offset angle $\Theta$. \\

&
& Ours
& Adjusts orientation so that an object's front faces the target, with $\Theta$ encoding preferred viewing or interaction directions (e.g., facing a desk, or seating area). \\
\midrule

\multirow{2}{*}[-0.4em]{Height-based} 

& $L_{\text{on top of}}(p_i, p_j, h)$
& \cellcolor{gray!10} LayoutVLM~\cite{sun2025layoutvlm}
& \cellcolor{gray!10} Position one asset on top of another. \\

&
& Ours
& Defines a vertical stacking relationship for placing smaller objects on surfaces while keeping sufficient interaction area. \\
\bottomrule

\end{tabular}

\footnotesize
\textbf{Notation:}
$p_i, p_j$ object poses;
$w_j$ wall index;
$b_i$ object bounding box;
$d_{\min}, d_{\max}$ minimum/maximum center distance;
$\Theta$ relative angle;
$h$ height offset.

\end{table*}

\subsubsection{VLM-based Sequential Group Optimization}
After the previous stages, the \textbf{[G] Sequential Group Optimization} stage compiles a constraint program that encodes the layout criteria and intra- and inter-group spatial relations as differentiable, anthropometrically grounded expressions. Each constraint incorporates the functional rationale and human behavioral conditions defined in the earlier stages, allowing the optimization process to preserve both semantic plausibility and operational validity. We represent all constraints as differentiable functions that enable the use of gradient-based optimization. Following the general optimization structure of LayoutVLM \cite{sun2025layoutvlm}, we formulated each violation as a continuous penalty for the object poses. However, unlike LayoutVLM, which derives its losses primarily from VLM-inferred spatial relations and physics-based collision terms, our violation terms are grounded in anthropometric reachability, functional clearance, and behavior-grounded spatial semantics. Specifically, we convert each inferred constraint into a differentiable penalty function:

\begin{itemize}
    \item \textbf{Distance constraints} produce violations when the center-to-center distance deviates from the inferred [$d_{\min}, d_{\max}$] range.
    \item \textbf{Orientation constraints} generate violations when object orientations differ from interaction-inferred alignment angles.
\end{itemize}

The optimization process minimizes the aggregated violation, $L = \sum_i w_i \cdot \text{violation}_i (\theta)$, where $i$ enumerates all behavior-aware, anthropometrically grounded constraints, \textbf{$w_i$} controls the relative influence of each constraint in the optimization, and $\text{violation}_i (\theta)$ computes a continuous penalty based on the deviation of object poses \textbf{$\theta$} from the constraint's target distance, orientation, or group-level configuration. We employ adaptive weighting where $w_i$ prioritizes collision avoidance when objects overlap significantly ($> 50\%$ bounding box overlap) while allowing flexible refinement of behavioral constraints. Optimization proceeds sequentially by group. For each group, we first minimize the intra-group constraint violations. As subsequent groups are optimized, inter-group constraints are applied to maintain behavioral relationships between previously placed groups and the current group. The optimizer runs for 400 gradient-based iterations, and the final layout is rendered in Blender using optimized object poses and floor-plan coordinates in meters.

\section{Technical Validation}
The following subsections describe the validation procedure, geometry based evaluation, user perception study, and the results that assess the layout quality of our approach relative to prior LLM-based methods.
\subsection{Validation Procedure}
In technical validation, our goal is to assess how the proposed scene generation framework, grounded in human-object interaction reasoning and anthropometric data, affects the layout quality relative to prior LLM-based approaches. We adopted a validation framework commonly used in 3D scene generation research, consisting of two components: \textbf{geometry-based metrics} (\textit{object-to-object} collision score and \textit{object-to-floor} in-boundary score) \cite{ccelen2024design, sun2025layoutvlm, sun2024haisor, hu2024mixeddiffusion3dindoor, feng2023layoutgpt} and a \textbf{user perception study} \cite{sun2025layoutvlm, sun2024haisor}.

\paragraph{Baseline}
We compared our method with a state-of-the-art system, LayoutVLM \cite{sun2025layoutvlm}. We selected the baseline because it is a representative LLM-based layout method that frames scene synthesis as a differentiable constraint optimization problem. The optimization structure enables a controlled comparison, in which the only major difference lies in how the constraints are obtained. The baseline relies on LLM common sense, whereas our method infers constraints from human-object interaction and anthropometric reference data. We did not include alternative LLM-based layout methods \cite{ccelen2024design, yang2024holodecklanguageguidedgeneration} because constraining them with fixed assets or deterministic placement would undermine their core generative process, precluding a fair comparison in controlled human-subject experiments where treatment consistency is necessary.

\paragraph{Setup.}
The validation covered five room types (Bedroom, Lounge, Office, Kitchen, and Dining) with two scenes each, totaling ten test scenarios with an average of 9.9 assets per room. All scenes were generated within a standardized 5.5 × 5.5 × 2.5m floor plan. For furniture assets, we selected 3D assets from the verified and preprocessed Objaverse \cite{deitke2023objaverse} for each room type following a prior methodology \cite{ccelen2024design, sun2025layoutvlm}. For each method and scene, we generated five candidate layouts using different random seeds and selected the layout with the highest collision-free score for analysis. This procedure yielded a matched set of 20 layouts (10 scenes $\times$ 2 methods), which we used for both the geometry-based evaluation in this section and the expert perception assessment.

\paragraph{Anthropometric data.}
Since our approach incorporates human dimensions into the constraint inference process, we utilized synthetic anthropometric profiles to validate this capability across diverse user bodies. First, we identified the 5\textsuperscript{th}--95\textsuperscript{th} percentile ranges for six key body dimensions (e.g., forward reach, lateral reach, body breadth, and body depth). We then provided the LLM with several example records sampled within these ranges and prompt it to generate a plausible combination of body dimensions that remains within the same percentile bounds. A synthetic profile was generated for each scene when the constraints were inferred, enabling technical validation to cover diverse users.

\subsection{Geometry-Based Evaluation}
To ensure a standardized quantitative comparison of physical validity, we adopted the widely used collision-free score and in-boundary score \cite{ccelen2024design, sun2025layoutvlm, sun2024haisor, hu2024mixeddiffusion3dindoor, feng2023layoutgpt}. These metrics were selected to verify the basic physical validity of the generated scenes, as physical feasibility (i.e., no overlaps, inside walls) is a necessary condition before a layout can be meaningfully evaluated for usability. For both metrics, higher values indicate greater physical plausibility. The \textbf{collision-free score} measures the proportion of object pairs that do not overlap:
\begin{equation}
    \text{CF} = \frac{1}{N_c} \sum_{(i,j)} \mathbb{I}(d_{ij} > r_i + r_j),
\end{equation}
where $d_{ij}$ is the distance between the object centers, $r_i$ and $r_j$ are the bounding-circle radii of objects $i$ and $j$, $\mathbb{I}(\cdot)$ returns 1 if the condition holds and 0 otherwise, and $N_c$ is the number of evaluated object pairs.
The \textbf{in-boundary score} captures whether all objects remain within the room:
\begin{equation}
    \text{IB} = \frac{1}{N_o} \sum_{i} \mathbb{I}(0 \le x_{\min}(i), x_{\max}(i) \le W, \ 0 \le y_{\min}(i), y_{\max}(i) \le D),
\end{equation}
where $x_{\min}(i), x_{\max}(i), y_{\min}(i),$ and $y_{\max}(i)$ denote the floor-plan bounding box extents of the object, $W$ and $D$ are the room width and depth, and $N_o$ is the number of objects.

\subsection{User perception study on the Generated Scene}
\paragraph{Participants}
We recruited 16 participants (six males and ten females; aged $M = 26.88, SD = 1.78$) who were professionals in interior design and architecture. All participants had completed a university degree in architecture or interior design, ensuring that they were familiar with spatial layout reasoning and furniture arrangements.

\begin{figure*}[t]
\centering
\includegraphics[width=\textwidth]{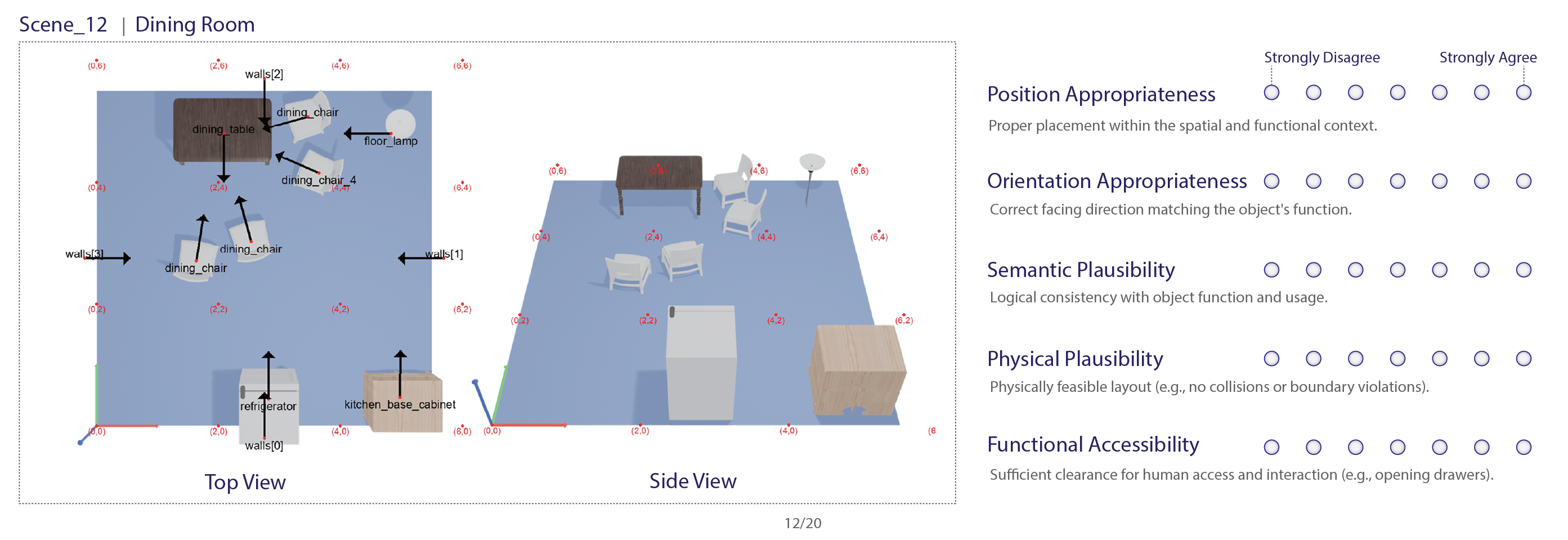}
\caption{Interface for User perception study. Red markers indicate floor-plan coordinates in meters, and arrows denote furniture orientations. Participants rated criteria on a 7-point Likert scale.}
\Description{A diagram shows the user perception study interface for a dining room scene (Scene_12). Top view and side view display the spatial layout with black arrows indicating furniture orientations. Participants rated five criteria on a 7-point Likert scale: Position Appropriateness evaluates proper placement within spatial and functional context, Orientation Appropriateness assesses object facing directions, Semantic Plausibility measures layout consistency with object function and usage, Physical Plausibility checks collision-free arrangements, and Functional Usability verifies sufficient clearance for human access and interaction.}
\label{fig:5}
\end{figure*}

\paragraph{Procedure.}
The study was conducted online. For each layout, the participants were shown two rendered images, a top view and a perspective view of the scene, and rated the layout on five criteria using a 7-point Likert scale (1 = Strongly Disagree, 7 = Strongly Agree). To support accurate judgment, the interface included orientation arrows and a metric floor plan (Figure~\ref{fig:5}). This resulted in 1,600 ratings ($16 \text{ participants} \times 20 \; \text{layouts} \times 5 \text{ criteria}$).

\paragraph{Measurements.}
We refined the evaluation criteria based on the baseline study \cite{sun2025layoutvlm} to capture the specific contributions of our method. The baseline originally evaluated layouts using Position, Orientation, and a combined physically grounded semantic alignment score (which aggregated physical plausibility and semantic consistency). However, a combined score limits the ability to distinguish between a layout that is simply logically correct versus one that is practically usable for a human. Since our approach emphasizes human-operational usability, we decomposed the baseline's composite metric into three granular criteria: semantic plausibility, physical plausibility, and functional usability. This separation allows us to isolate and measure how our anthropometric grounded constraints improve the detailed operability of the scene. Participants evaluated their agreement with statements regarding:

\begin{itemize}
    \item \textbf{Position appropriateness:} whether each object is placed at an appropriate location for its function and use within the room (e.g., not unreasonably far from related furniture).
    \item \textbf{Orientation appropriateness:} whether each object faces a direction that matches its function (e.g., a chair facing a desk or TV facing the seating area).
    \item \textbf{Semantic plausibility:} whether the arrangement of furniture is logically consistent with the intended use of the space (e.g., a bed placed meaningfully relative to a bedside table).
    \item \textbf{Physical plausibility:} whether the layout appears physically feasible in terms of basic spatial rules (e.g., no obvious collisions or boundary violations).
    \item \textbf{Functional usability:} whether people seem able to approach, access, and operate each object in a realistic way (e.g., sufficient space to open drawers fully, pull out chairs, or stand and work on a surface).
\end{itemize}

\subsection{Results of Technical Validation}

\paragraph{Geometry-Based Evaluation Results.}
Regarding the comparison of physical validity, our method achieved a collision-free score of 90.4 and an in-boundary score of 73.4, compared to the baseline's scores of 92.2 and 63.2. Although our framework showed a slightly lower collision-free score ($-1.8\%$), it demonstrated substantially improved boundary adherence ($+10.2\%$). Collision-free and in-boundary score metrics provide objective, geometry-based indicators of physical plausibility by measuring whether objects overlap and whether assets remain within room boundaries.

\begin{table}[t]
\centering
\small
\renewcommand{\arraystretch}{1.5}
\caption{Results of the user perception study. Values represent the median, with the interquartile range [1\textsuperscript{st} quartile, 3\textsuperscript{rd} quartile] shown in brackets. Statistical significance was determined using the Wilcoxon Signed-Rank test with Holm-Bonferroni correction (*\textit{p} < 0.01).}
\label{tab:user_perception}
\begin{tabular}{@{\hspace{0.5em}}lcccc@{\hspace{0.5em}}}
\toprule
\textbf{Measurements} & \textbf{Baseline} & \textbf{Ours} & $\mathbf{p_{adj}}$ & \textbf{r} \\
\midrule
\rowcolor{gray!10} Position appropriateness    & \makecell{3.00\\{\scriptsize[2.25, 4.00]}} & \makecell{\textbf{5.00}\\{\scriptsize[4.38, 6.00]}} & $<$0.01 & 0.85 \\
Orientation appropriateness & \makecell{3.00\\{\scriptsize[2.00, 3.25]}} & \makecell{\textbf{5.00}\\{\scriptsize[4.50, 6.13]}} & $<$0.01 & 0.88 \\
\rowcolor{gray!10} Semantic plausibility       & \makecell{3.00\\{\scriptsize[2.75, 3.63]}} & \makecell{\textbf{5.00}\\{\scriptsize[4.38, 6.00]}} & $<$0.01 & 0.86 \\
Physical plausibility       & \makecell{3.00\\{\scriptsize[2.00, 4.13]}} & \makecell{\textbf{5.00}\\{\scriptsize[5.00, 6.00]}} & $<$0.01 & 0.78 \\
\rowcolor{gray!10} Functional usability        & \makecell{3.00\\{\scriptsize[2.00, 4.00]}} & \makecell{\textbf{5.00}\\{\scriptsize[4.50, 6.00]}} & $<$0.01 & 0.77 \\
\bottomrule
\end{tabular}
\end{table}

\paragraph{User Perception Study Results}
When examining the results of the user perception study, our method consistently outperformed the baseline across all five evaluation criteria (Table~\ref{tab:user_perception}). Our method achieved a median of 5.0 in all metrics, whereas the baseline remained at a median of 3.0. Looking at the distribution, our method achieved 1\textsuperscript{st} quartile values ranging from 4.375 to 5.0 and 3\textsuperscript{rd} quartile values from 6.0 to 6.125. In contrast, the baseline showed significantly lower performance, with 1\textsuperscript{st} quartile values between 2.0 and 2.75, and 3\textsuperscript{rd} quartile values between 3.25 and 4.125. Notably, while the baseline achieved a higher collision-free score, this difference did not translate into improved user perception. Statistical analysis using the Wilcoxon Signed-Rank test with Holm-Bonferroni correction confirmed that these differences were statistically significant across all five criteria (adjusted $p < 0.01$). Furthermore, the effect sizes were large ($r > 0.77$). These results indicate that experts recognize our layouts as substantially more suitable for human operation than the baseline.

\begin{figure*}[t]
\centering
\includegraphics[width=\textwidth]{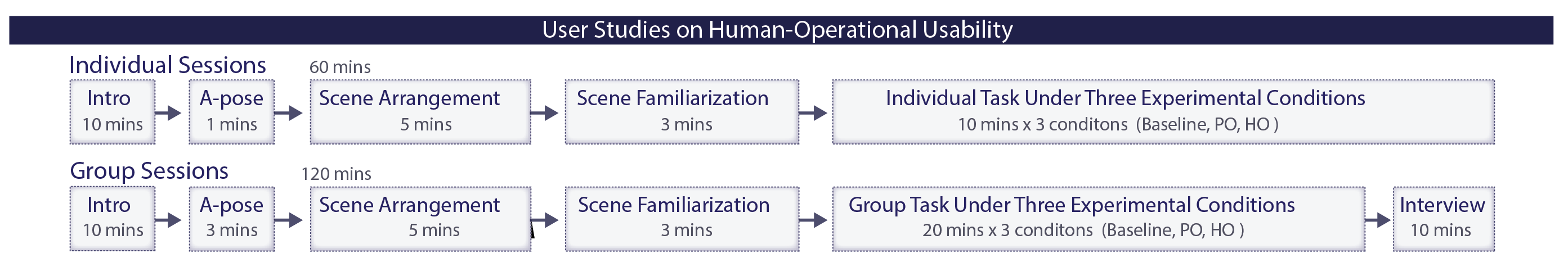}
\caption{\textbf{Experimental Procedure Flowcharts for Individual and Group Sessions.}}
\Description{A flowchart shows the experimental procedure for user studies. Individual Sessions last 60 minutes with phases including intro, A-pose capture, scene arrangement, familiarization, and task performance under three conditions. Group Sessions last 120 minutes following a similar structure with an additional interview phase at the end.}
\label{fig:6}
\end{figure*}

\paragraph{Limitations and Motivation for Usability Study.}
Geometry-based evaluation metrics have limitations when assessing human usability. First, they evaluate scenes under fixed-object assumptions (e.g., non-articulated drawers), and therefore cannot consider dynamic object functionality or actual human behavioral patterns. Second, while the user perception study complements these metrics by incorporating expert judgment, it relies solely on visual estimation from static renderings. Evaluating a layout visually does not guarantee that it supports actual physical operations---such as comfortable reaching or unhindered movement---especially for users with diverse body dimensions. This limitation motivated the human-operational usability study, which is a task-based user study to evaluate the generated scenes in realistic usage scenarios.

\section{User Studies on Human-Operational Usability}

To investigate whether our behavior-aware and anthropometric approach translates into actual operational benefits, we conducted a human-centered usability evaluation based on layouts parameterized by each participant's measured anthropometric data. We designed two complementary user studies to examine how participants physically moved and acted within these scenes from different perspectives. The experimental procedure flow for both sessions is illustrated in Figure~\ref{fig:6}.

\begin{itemize}
    \item \textbf{Individual sessions} ($N=20$): Provides controlled, quantitative measures of how anthropometrically grounded, behavior-aware layouts support task performance. Through structured single-user tasks, we captured movement behaviors, quantified task efficiency and interaction-space usage, and obtained data that enabled the statistical assessment of our approach.
    \item \textbf{Group sessions} ($N=18$, six teams of three): Evaluates the same layouts in realistic multi-user scenarios where multiple occupants naturally share space. By analyzing cumulative movement trajectories and assessing post-task interview feedback, we captured how layouts support spatial negotiation and collaborative activities, which are conditions that reflect real-world shared environments.
\end{itemize}

\subsection{Setup for Individual and Group Studies}
The following experimental conditions, environments, apparatus, and data collection procedures were applied consistently across both the individual and group studies.

\subsubsection{Experimental Conditions}
\begin{figure*}[t]
\centering
\includegraphics[width=0.90\textwidth]{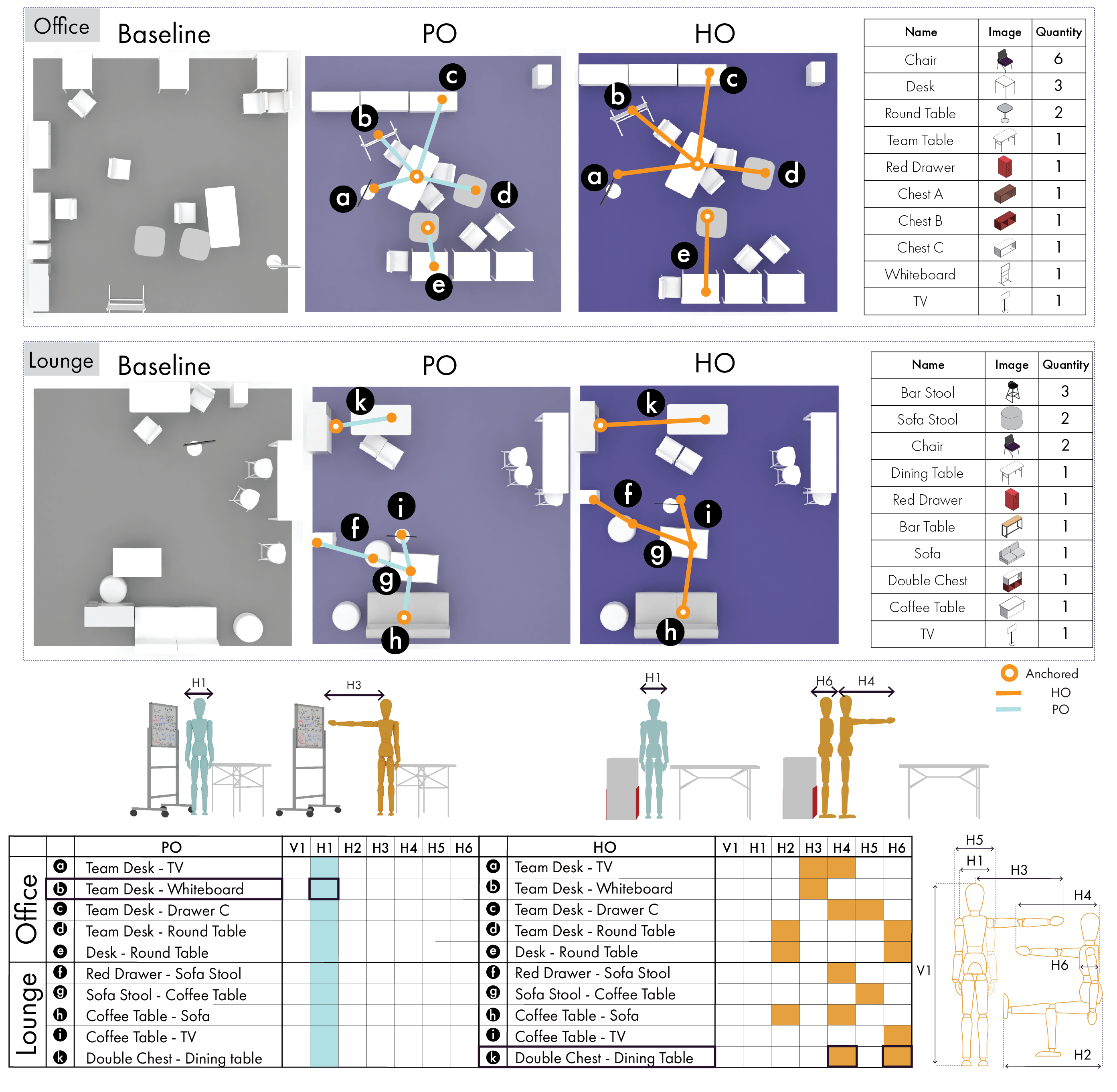}
\caption{\textbf{Baseline, PO, and HO layouts across office and lounge environments.}}
\Description{A diagram compares three layout conditions (Baseline, PO, HO) across office and lounge settings. Floor plans show furniture arrangements with orange lines indicating where anthropometric constraints are applied between furniture pairs. Asset inventories list furniture items with thumbnails and quantities. The lower section maps inter-object relations to anthropometric measures, showing horizontal body dimension parameters (H1-H6, V1) used to calculate spatial distances between furniture pairs.}
\label{fig:7}
\end{figure*}

We evaluated three conditions: a geometry-based \textbf{baseline \cite{sun2025layoutvlm}} and two versions of our framework, \textbf{Passage-Only (PO)} and \textbf{Human-Operational (HO)}, which vary only in the type of anthropometric parameters injected into spatial constraints (Figure~\ref{fig:7}).

\begin{itemize}
    \item\textbf{Baseline:} Layouts generated without any personalized anthropometric information; spatial relations are determined solely by generic geometric constraints and collision checks.
    \item\textbf{PO:} Our framework with \textit{static anthropometric dimensions (e.g., body width, torso depth)} applied to guarantee minimal passage width and basic navigability.
    \item\textbf{HO:} Our framework with \textit{movement-based anthropometric dimensions (e.g., extended arm reach, forward reach, seating buttock-to-toe length)} applied to ensure adequate operational space for human-object interactions.
\end{itemize}

PO and HO share the same spatial relationship and constraint types; only the minimum and maximum distances differ: static body sizes in PO versus movement envelopes in HO. This design enables two key comparisons: comparing PO and HO isolates the effect of dynamic, task-aware anthropometrics, because any usability differences directly reflect the added value of modeling movement-based operational space. Comparing baseline and HO assesses the full benefit of our framework, which integrates semantic scene understanding, spatial relation inference, and personalized anthropometric parameterization. This comparison highlights how behavior-aware layouts improve upon geometry-based methods by aligning spatial constraints with actual human movement.

\subsubsection{Evaluation Environments}
We selected office and lounge environments as representative indoor spaces, where diverse object arrangements and human activities occur naturally, supporting work, collaboration, storage, navigation, and rest tasks, while maintaining sufficient experimental control. All three conditions (Baseline, PO, HO) used the same 5.5 × 5.5 × 2.5m floor plan and were generated following the same procedure used in the technical validation. For the user studies, the PO and HO were instantiated on a per-participant basis by substituting each participant's measured anthropometric profile into the distance constraints, whereas the baseline remained purely geometry-based. Across all conditions, the office and lounge scenes shared the same furniture sets and overall scene topology (Figure~\ref{fig:7}).

\subsubsection{Motion Capture and Apparatus}
\begin{figure*}[t]
\centering
\includegraphics[width=0.90\textwidth]{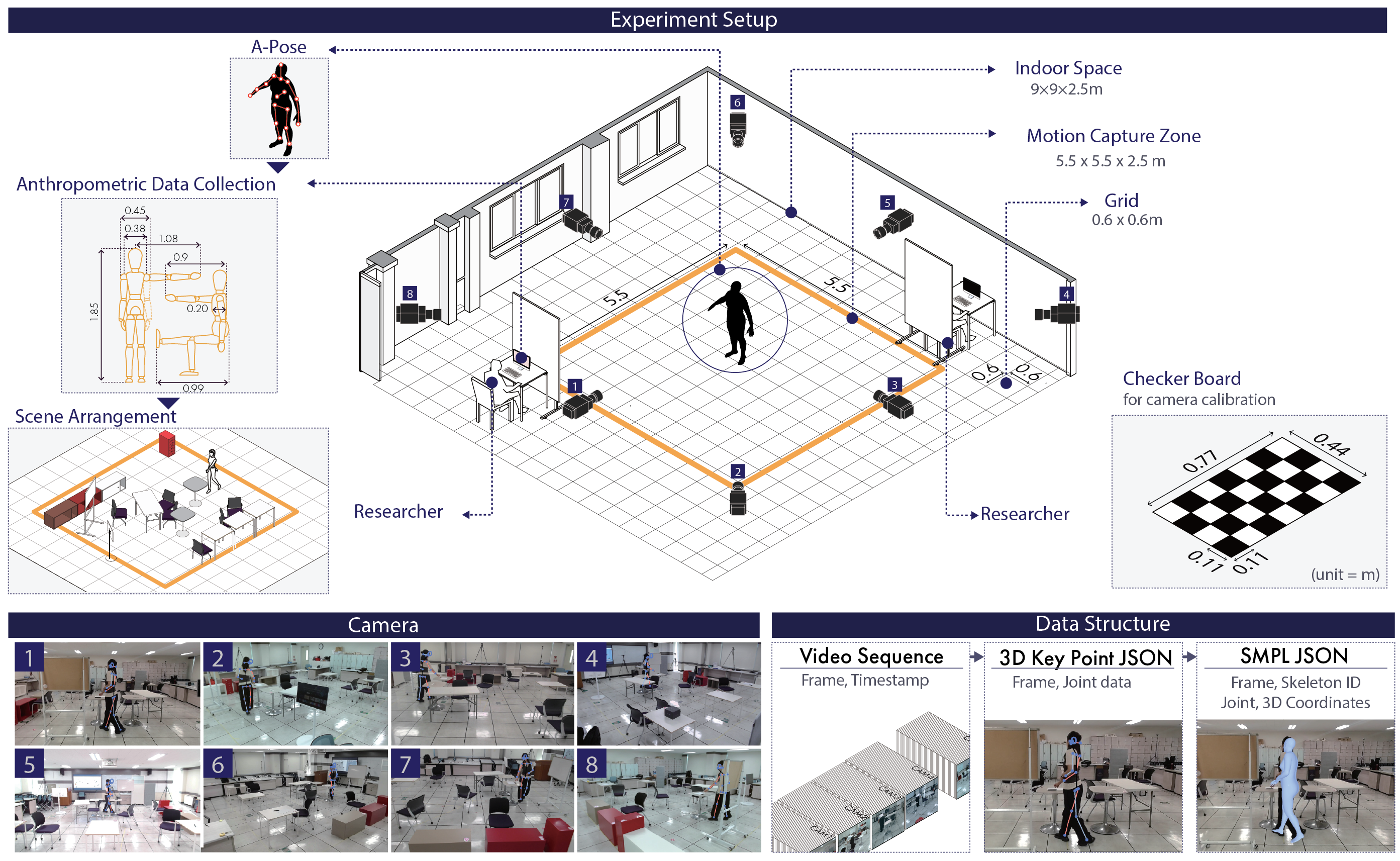}
\caption{\textbf{Experimental environment, motion capture setup, and data processing pipeline.} The illustration depicts the physical configuration of the 8-camera array and capture zone, alongside the sequential data flow from raw video to 3D SMPL model reconstruction. All spatial measurements are in meters.}
\Description{A diagram shows the experimental environment, motion capture setup, and data processing pipeline. The physical space configuration includes an indoor space measuring 9x9x2.5m, a motion capture zone of 5.5x5.5x2.5m, an 8-camera array positioned around the perimeter, a grid floor of 0.6x0.6m, and a checkerboard for camera calibration. Anthropometric data collection shows A-pose capture with SMPL body landmarks. The camera section shows 8 synchronized views capturing participant movements. The data structure section illustrates the processing pipeline from video sequences to 3D keypoint JSON and SMPL JSON, containing frame timestamps, joint data, and skeleton coordinates.}
\label{fig:8}
\end{figure*}

We employed a markerless motion capture setup using EasyMocap \cite{shuai2022multinb} to obtain absolute human coordinates and body measurements for both scene generation and human-scene interaction evaluation. We installed eight RGB cameras (Logitech webcams) in a $9 \times 9 \times 2.5$m indoor space and performed intrinsic and extrinsic calibration using a $4 \times 7$ checkerboard \cite{shuai2022multinb}. The calibrated cameras recorded synchronized multi-view videos at 4K/60fps within a capture volume of $5.5 \times 5.5 \times 2.5$m. To reconstruct the 3D body motion, the recorded data were processed frame-by-frame through (1) 2D keypoint extraction, (2) 3D triangulation, and (3) SMPL fitting. The experimental setup, camera placement, and captured data types are shown in Figure~\ref{fig:8}.

\subsubsection{Anthropometric Data Collection}
Participant-specific anthropometric measurements were extracted from the fitted SMPL models to parameterize the distance constraints used in the PO and HO conditions. Accordingly, we adopted the pose-independent 3D anthropometry method of Bojani{\'c} et al. \cite{bojanic2024pose}, which predicts standardized body measurements from sparse 3D landmarks. We selected this method because it accurately reconstructs canonical A-pose dimensions regardless of the participant's actual pose during capture. This eliminates the need for strict A-pose scans while maintaining accuracy comparable to dense-geometry approaches. From each participant's reconstructed SMPL body, we first extracted the 3D coordinates of the 70 body landmarks, and passed them through the pre-trained prediction model to obtain a set of standardized anthropometric dimensions. For additional task-specific operational dimensions not directly provided by the Bojani{\'c} model, such as specific reach distances or joint-to-joint separations relevant to our interaction scenarios, we computed the Euclidean distances between the selected pairs of SMPL landmarks. The resulting combined profile served as the quantitative basis for instantiating personalized object--object distance constraints.

\begin{table*}[t]
\centering
\small
\caption{Sequential task list for office and lounge environments.}
\label{tab:tasks}
\renewcommand{\arraystretch}{1.2}
\begin{tabular}{@{\hspace{1em}}p{0.45\textwidth}@{\hspace{2em}}p{0.45\textwidth}@{\hspace{1em}}}
\toprule
\textbf{Task Sequence for Office} & \textbf{Task Sequence for Lounge} \\
\midrule
\rowcolor{gray!10} O1. Take box out of Chest C, place it on top. & L1. Move to Sofa Stool 1, sit down. \\
O2. Move box onto Desk C, sit in front of it. & L2. Take cup from Coffee Table, move to Sofa Stool 2, sit down. \\
\rowcolor{gray!10} O3. Stand up, take pen from third compartment on Red Drawer. & L3. Holding cup, move to Bar Table, sit down. \\
O4. Move to Round Table 2, use remote control to turn on TV. & L4. Holding cup, walk to Double Chest, place cup inside compartment. \\
\rowcolor{gray!10} O5. On Round Table 1, write two content names from TV onto paper. & L5. Find remote control in Red Drawer, move to Sofa, sit down. \\
O6. Holding paper, move to Chest B, take out board marker. & L6. From Sofa, turn on TV, find content, read title aloud. \\
\rowcolor{gray!10} O7. Holding paper and marker, move to Team Table, sit down. & L7. Place remote control on Coffee Table. \\
O8. Move to Whiteboard, copy the two words you wrote. & L8. Move to Double Chest, take out pen and paper. \\
\rowcolor{gray!10} O9. Sit at Team Table, read Whiteboard words backward from right. & L9. Move to Dining Table, sit, write content title on paper. \\
O10. Move to Desk A, sit down, write remembered words onto paper. & L10. Stand up, place paper and pen in lower compartment of Double Chest. \\
\bottomrule
\end{tabular}
\end{table*}

\section{User Study 1: Individual Sessions}
Individual sessions examined how the three scene conditions--Baseline, PO, and HO--affect the task-level usability for a single user in personalized layouts. Building on anthropometric personalization and motion capture framework, this study focuses on a controlled, quantitative evaluation of human–operational performance. Specifically, we compared three conditions in terms of task completion time, trajectory efficiency during navigation, and interaction-space utilization around key objects.

\subsection{Experimental Setup}
\paragraph{Participants.}
The individual sessions involved 20 participants: 10 in the office space (four males and six females; aged $M = 24, SD = 1.89$) and 10 in the lounge space (five males and five females; aged $M = 24.3, SD = 2.63$).

\begin{figure*}[t]
\centering
\includegraphics[width=\textwidth]{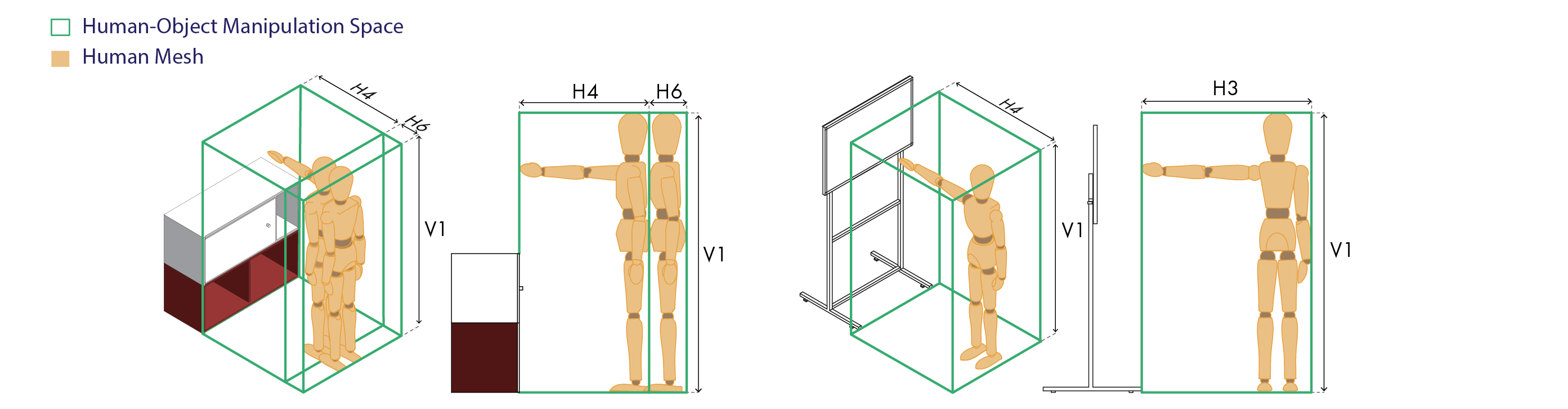}
\caption{\textbf{Definition of Human-Object Manipulation Space.} The green wireframe defines the Human-Object Manipulation Space, a bounding volume parameterized by the participant’s specific anthropometric dimensions (e.g., vertical height V1, horizontal reach H3, H4).}
\Description{A diagram defines Human-Object Manipulation Space as a 3D bounding box (green wireframe) around target objects, parameterized by participant-specific anthropometric dimensions. The visualization shows the human mesh (orange) within the manipulation space across multiple viewpoints. Participant height defines the vertical extent, while anthropometric distance constraints determine width and depth}
\label{fig:9}
\end{figure*}

\paragraph{Procedure and Task.}
At the beginning of each session, we explained the study purpose and procedure and then captured each participant in an A-pose in an empty space using the motion capture setup. We fitted the captured data to the SMPL models, computed anthropometric measurements, and used these measurements to instantiate the PO and HO layouts for each participant. Baseline layouts were fixed for each environment. While one researcher provided task instructions and a short practice session, the others prepared the physical environment by arranging furniture according to the designated layout. The order of the three layout conditions (Baseline, PO, HO) was counterbalanced by using a Latin square. The participants received safety instructions and an explanation of furniture naming conventions before the main trials. For each assigned environment (office or lounge), we defined ten tasks that reproduced everyday situations, such as navigating between furniture, sitting, and opening drawers, to ensure that all furniture items were used at least once (Table~\ref{tab:tasks}). Task instructions followed the \textit{object--action--target} format \cite{wang2022humanise, achlioptas2020referit3d}. To preserve the initial layout, the participants were instructed not to move furniture except for chairs. The tasks were connected sequentially; the endpoint of one task served as the starting point for the next task. Each participant executed ten tasks in their assigned space under all three conditions, resulting in 60 recorded interaction sequences ($20 \text{ participants} \times 3 \text{ conditions}$), with each session lasting approximately 60 minutes.

\subsection{Measurements and Data Analysis}
Our measurement set was designed to evaluate whether the proposed anthropometric constraints yield measurable usability improvements. Each metric links the imposed anthropometric constraints to a specific aspect of use: task performance, trajectory efficiency, or interaction-space utilization. We focused on four metrics: 

\paragraph{Task Completion Time.} We measured the task completion time from the moment the participants began to act after receiving the full task instructions to the moment they verbally indicated completion. Three researchers independently reviewed all recordings at the frame level to annotate precise start and end points, and cross-checked their annotations to ensure consistency.

\paragraph{Trajectory Counting.} To analyze how layouts shape movement paths, we projected the SMPL pelvis joint onto a 2D plane to generate time-series trajectories and then counted the number of distinct paths taken. Trajectories were treated as distinct when they involved different detours around furniture or obstacles, thereby quantifying the number of different paths that participants used to complete identical tasks.

\paragraph{Sequence Action Labeling.} To capture the action-level differences in movement behavior, we used MMAction2 \cite{2020mmaction2}, a spatio-temporal action detection model, to automatically label the participant action sequences. We divided each video into short 6-frame segments and detected the actions for each segment. For every 6-frame segment in each of the three camera views, we extracted the top two action labels with confidence scores greater than 0.6. Then, we combined the labels from the three views and weighed them by confidence to obtain the final action sequence for each task.

\paragraph{Volumetric Occupancy Ratio.} This metric measures the extent to which the intended interaction-space around an object is used by the participant. We first defined a human-object manipulation space as a 3D bounding box around each target object based on object-specific interaction distances derived from each participant's anthropometric profile. The participant's height was set to the vertical extent, and the anthropometric distance constraints determined the width and depth (Figure~\ref{fig:9}). We focus on objects with movable parts that require sufficient clearance: a double chest, a single chest, a whiteboard in the office, and a red drawer in the lounge. Using Blender and a Python script, we computed (1) the accumulated 3D volume occupied by the participant's body during task execution (the temporal union of human mesh volumes across all frames) and (2) its intersection with the human-object manipulation space. 

$M$ denotes the cumulative human mesh volume (temporal union across frames) and $B$ denotes the
defined Human-Object Manipulation Space (bounding box). The ratio is defined as:

\[
\text{Volumetric Occupancy Ratio} = \frac{\text{Vol}(M \cap B)}{\text{Vol}(B)}
\]

where \(\text{Vol}(\cdot)\) represents the volume operator and \(\cap\) denotes the intersection.
We computed this ratio for each task and averaged the results to obtain a final metric per participant.

Experimental outcomes were analyzed using mixed design analyses of variance (ANOVAs) with layout (Baseline, PO, HO) as a within-subjects factor and environment (office, lounge) as a between-subjects factor. For each metric, we first tested the layout × environment interaction to verify consistent effects across environments, then examined the Layout main effect. Given the modest sample size, follow-up pairwise comparisons used Wilcoxon signed-rank tests with Bonferroni correction ($\alpha = 0.017$) as a conservative approach.

\subsection{Results of User Study 1: Individual Sessions}
Mixed-design ANOVAs revealed significant main effects of layout for task completion time ($F$(2,36) = 4.38, $p$ = 0.020), trajectory counting ($F$(2,36) = 7.45, $p$ = 0.002), sequence action labeling ($F$(2,36) = 32.49, $p$ < 0.001), and volumetric occupancy ratio ($F$(2,36) = 66.12, $p$ < 0.001). No significant layout × environment interactions were observed (all $p$s > 0.05), indicating consistent effects across both environments. Below, we report detailed comparisons for each metric.

\subsubsection{Task Completion Time}
\begin{figure*}[t]
\centering
\includegraphics[width=0.95\textwidth]{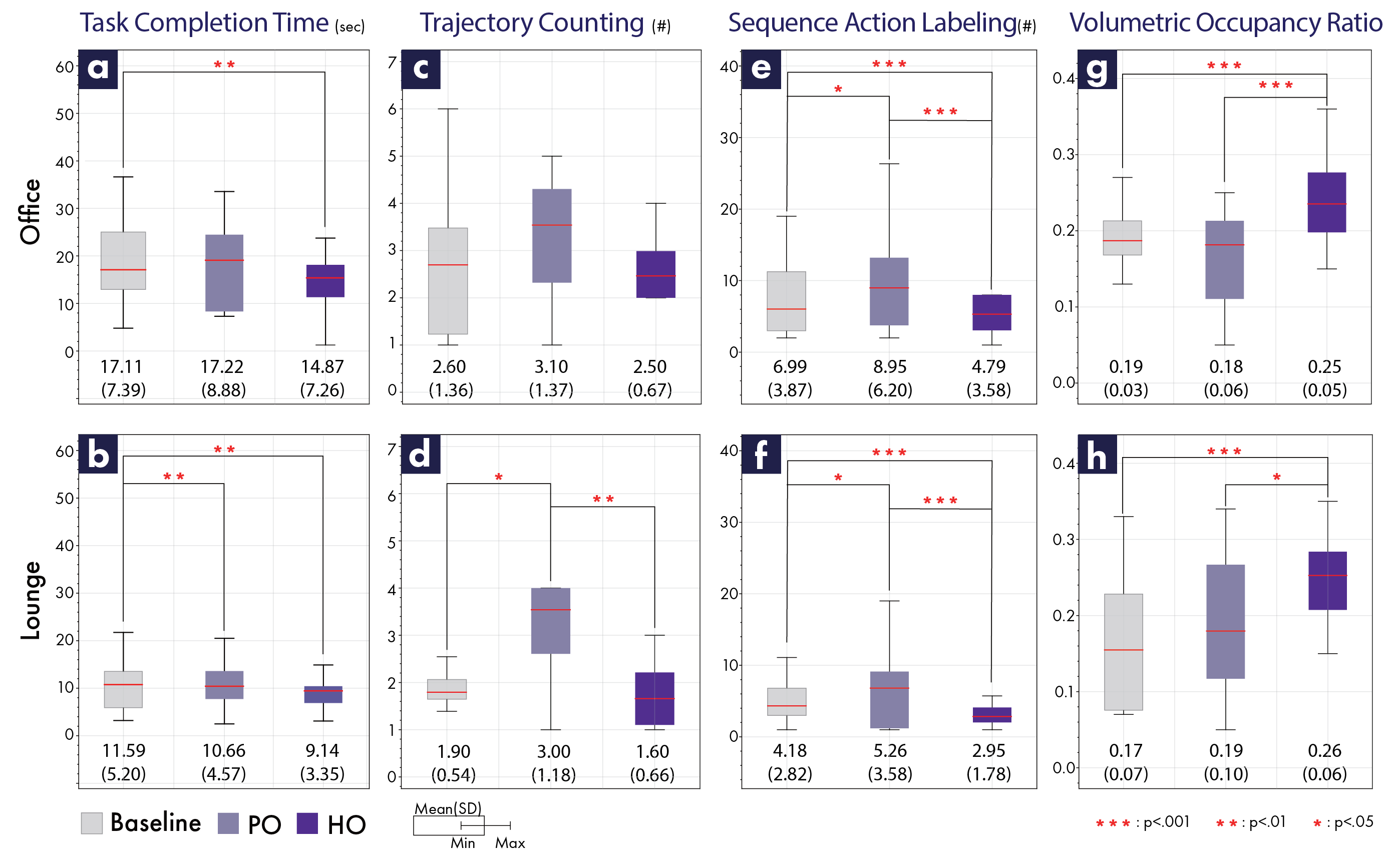}
\caption{Results of Measurements. (a, b): Task Completion Time.  (c, d) : Trajectory Counting. (e, f) : Sequence Action Labeling.  (g, h) : Volumetric Occupancy Ratio.}
\Description{A diagram shows measurement results across four metrics comparing three layout conditions (Baseline, PO, HO) in office and lounge settings. Box plots display Task Completion Time in seconds (a, b), Trajectory Counting as number of distinct paths (c, d), Sequence Action Labeling as action counts (e, f), and Volumetric Occupancy Ratio as spatial utilization proportion (g, h). Results show HO consistently achieves shorter task times, fewer trajectories, reduced action counts, and higher occupancy ratios compared to baseline and PO conditions across both environments.
}
\label{fig:10}
\end{figure*}

A lower task completion time indicates a more efficient layout. When participants perform the same sequence of tasks, faster completion implies that the layout allows shorter or less obstructed routes and easier access to target objects. Therefore, we compared the total time required to complete ten sequential tasks in an individual study, where each task endpoint served as the starting point for the next task. In the office environment, HO achieved a mean completion time of 14.87 seconds, compared with 17.11 seconds for the baseline and 17.22 seconds for PO (Figure~\ref{fig:10}a). The difference between baseline and HO was statistically significant ($p < 0.01$). HO achieved the shortest completion time, requiring $13.1\%$ less time than baseline. Similar patterns were observed in the lounge environment, where HO required 9.14 seconds, PO required 10.66 seconds, and baseline required 11.59 seconds (Figure~\ref{fig:10}b). The task completion time differences in the lounge showed significant differences between baseline and HO ($p < 0.01$) with $21.1\%$ reduction compared to baseline.

\subsubsection{Trajectory Efficiency}
We measured the layout trajectory efficiency using trajectory counting and sequence action labeling to assess whether the layouts induced intuitive path selection and to analyze the economic efficiency through the number of trajectories used.

\begin{figure*}[t]
\centering
\includegraphics[width=0.9\textwidth]{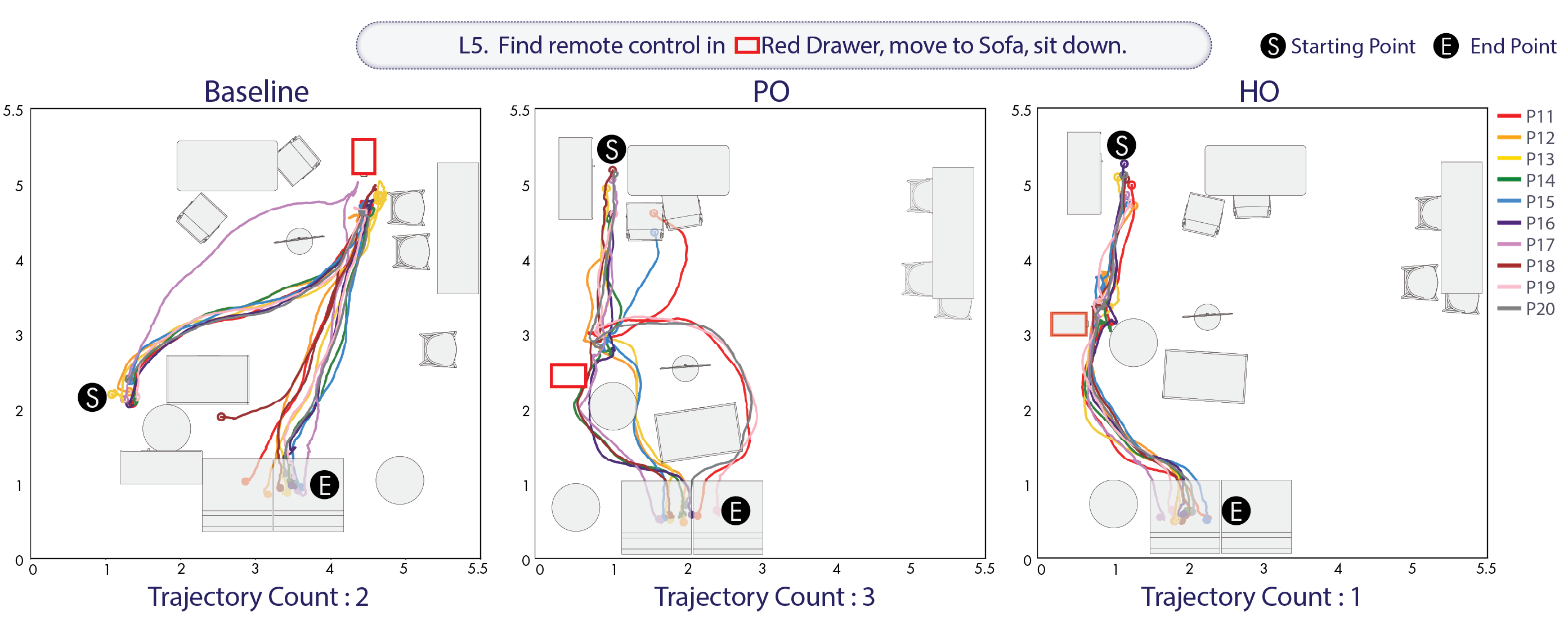}
\caption{\textbf{Examples of trajectory counting and participant trajectory visualization for L5.}}
\Description{A diagram illustrates trajectory counting methodology and multi-participant trajectory visualization for lounge task “L5: Find the remote control in the red drawer, move to the sofa, sit down.” Floor plans show participant movement paths from starting point (S) to end point (E) across three layout conditions. Baseline condition shows detoured paths around furniture. PO condition displays three trajectories with detours caused by furniture obstacles. HO condition demonstrates one consistent trajectory with direct, unobstructed paths. Each participant's path is color-coded (P11-P20).
}
\label{fig:11}
\end{figure*}

\paragraph{Trajectory Counting.}
A lower trajectory count indicates a more intuitive and efficient layout (Figure~\ref{fig:11}). If the count is 1, all participants who experienced that layout followed the same path, suggesting that the scene allowed a clear and natural route. The mean trajectory counts across all sequences showed a consistent pattern of HO < baseline < PO in both office and lounge layouts, with HO demonstrating the lowest trajectory count. In the office, HO generated an average of 2.5 distinct paths per task across all participants, the baseline generated 2.6 paths, and PO generated 3.1 paths, although these differences were not statistically significant (Figure~\ref{fig:10}c). In the lounge, HO showed 1.6 paths, baseline showed 1.9 paths, and PO generated 3.0 paths (Figure~\ref{fig:10}d). HO required fewer paths compared to PO ($p < 0.01$) with a $46.7\%$ reduction, and PO showed more paths than baseline ($p < 0.05$).

\paragraph{Sequence Action Labeling.}
We counted frame-by-frame labeled actions to determine the action sequences required for task completion, and calculated the mean action counts per task. A lower action count indicates a more efficient trajectory; additional actions typically arise from unnecessary behaviors, such as moving obstacles or making corrective adjustments. For example, in task L4, participants were required to stand up from the bar table while holding a cup, walk to the double chest, and place the cup inside a compartment. In the HO lounge layout, P12 completed the task using four actions \{sit-hold-walk-stand\}. In the baseline layout, an extra `bend' action was required to move an obstructing stool, resulting in \{sit-hold-walk-stand-bend-stand\}. Therefore, fewer actions reflected a more direct and efficient trajectory for task completion. In office layouts, HO required an average of 4.79 actions, whereas baseline required 6.99 actions and PO required 8.95 actions (Figure~\ref{fig:10}e). HO required significantly fewer actions than both baseline ($p < 0.05$) with $31.5\%$ reduction and PO ($p < 0.01$) with $46.5\%$ reduction. In lounge layouts, HO required an average of 2.95 actions, baseline required 4.18 actions, and PO required 5.26 actions (Figure~\ref{fig:10}f). HO showed significantly lower action counts than both baseline ($p < 0.001$) with $29.4\%$ reduction and PO ($p < 0.001$) with $43.9\%$ reduction.

\subsubsection{Interaction-Space Utilization}
\begin{figure*}[t]
\centering
\includegraphics[width=0.90\textwidth]{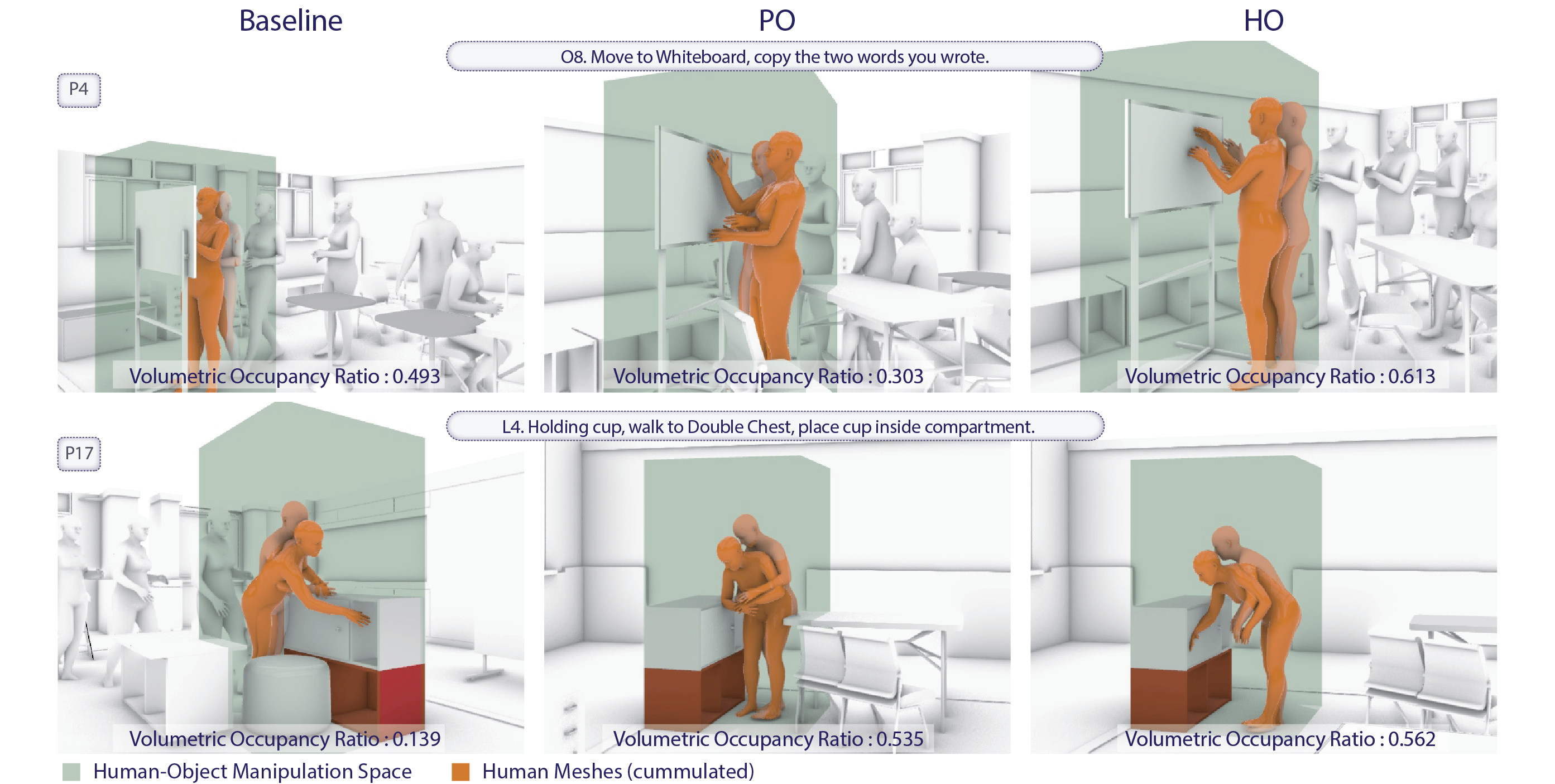}
\caption{\textbf{Graphical Examples of Human-Object Manipulation Space (green box) and SMPL meshes for P4 (O8) and P17 (L4).}}
\Description{A diagram shows graphical examples of Human-Object Manipulation Space (green wireframe box) and accumulated human meshes (orange) for P4 performing office task O8 and P17 performing lounge task L4. For O8 (Move to the Whiteboard, copy the two words you just wrote down), baseline shows a Volumetric Occupancy Ratio of 0.493 with partial space utilization, PO shows 0.303 with restricted lateral movement due to furniture proximity, and HO shows 0.613 with full frontal access and efficient space usage. For L4 (Holding the cup, stand up, walk to the Double Chest, place the cup inside the compartment with a door), baseline shows 0.139 with awkward side approach, PO shows 0.535 with constrained posture, and HO shows 0.562 with decent interaction enabled by adequate clearance.}
\label{fig:12}
\end{figure*}

For the Volumetric Occupancy Ratio, higher values indicated better interaction-space utilization; participants occupied more of the intended human-object manipulation space and could operate objects more smoothly. Qualitative examples illustrate this pattern (Figure~\ref{fig:12}). In the PO office layout, P4 could only use a part of the Whiteboard because the Team Table was placed too close; therefore, the accumulated body volume within the Whiteboard's manipulation box remained low. In the baseline lounge layout, P17 had to work sideways at the Double Chest because a Sofa Stool blocked the front access area, again yielding sparse occupancy within the manipulation space. In contrast, in both scenarios, the HO layouts showed dense, evenly distributed occupancy within the manipulation boxes while participants performed the associated tasks. In the office layouts, HO showed an average occupancy ratio of 0.25, PO showed 0.18, and baseline showed 0.19 (Figure~\ref{fig:10}g). The lounge layouts showed similar results (Figure~\ref{fig:10}h). In office layouts, HO showed significantly higher occupancy ratios than both baseline ($p < 0.001$) with $31.6\%$ increase and PO ($p < 0.001$) with $38.9\%$ increase. Lounge layouts showed HO significantly outperforming baseline ($p < 0.001$) with $52.9\%$ increase, and PO showing higher ratios than baseline ($p < 0.01$) with $11.8\%$ increase.

\subsubsection{Summary}
Across all four metrics, the HO condition consistently provided better task-level usability than the baseline and PO layouts.
\begin{itemize}
    \item\textbf{Task Completion Time:} HO demonstrated the shortest completion times in both the office and lounge environments, whereas PO showed no improvement over baseline.
    \item \textbf{Trajectory Counting:} The mean trajectory counts followed a consistent pattern of HO < baseline < PO, indicating that HO layouts induced more intuitive routes, whereas PO often led to additional detours.
    \item \textbf{Sequence Action Labeling:} HO required the fewest actions per task in both environments, reducing extra behaviors such as moving obstacles or making corrective adjustments compared to the baseline and PO.
    \item \textbf{Volumetric Occupancy Ratio:} HO achieved the highest interaction-space utilization, whereas PO and baseline left more of the intended manipulation space unused.
\end{itemize}

\section{User Study 2: Group Sessions}
Although individual sessions provide controlled, single-user measurements of task-level performance, many real-world environments are occupied by multiple people who perform overlapping activities over extended periods. To account for these multi-user dynamics, where layout usability depends not only on individual route efficiency but also on how occupants share space, experience mutual interference, and adapt their movements in response to others, we conducted group sessions in which small teams performed continuous collaborative work in the same baseline, PO, and HO layouts used in the individual study. In addition to the motion capture data, we conducted post-interviews to qualitatively examine perceived crowding, coordination, and workflow support in each layout condition.

\subsection{Experimental Setup}
\paragraph{Participants.}
The group sessions involved 18 participants organized into six teams. In the office space, three teams (nine participants: three males and six females; aged $M = 23.56$, $SD = 2.92$) performed tasks, whereas the remaining three teams (nine participants: four males and five females; aged $M = 23.78$, $SD = 2.49$) performed tasks in the lounge space.

\paragraph{Procedure and Tasks.}
At the start of each session, the teams were briefed on the study and recorded using the same motion capture and anthropometric framework. For scene generation, we constructed a team-level anthropometric profile by taking the maximum value among the three team members for each dimension and used this profile to instantiate the PO and HO constraints. After safety instructions and a short practice session, we arranged the furniture for the current condition and asked teams to perform a series of collaborative coloring and sorting tasks in each of the three layouts (Baseline, PO, HO), following spatial-interaction protocols adapted from \cite{luo2022should, luo2025documents}. In each layout, participants viewed templates on the TV, retrieved materials from storage units, worked at individual desks using hints from the shared space, reconvened to discuss the sorting criteria shown on the TV, and attached the final templates to the whiteboard. The task materials were distributed across the space to encourage resource exchange and movement, naturally inducing circulation, path intersections, and clustering. After all three conditions, we conducted individual post-interviews in a separate room. Each team session lasted approximately 120 minutes, yielding 18 group interaction sequences in total (six teams $\times$ three conditions).

\subsection{Data Analysis and Visualization}
\paragraph{Spatial Usage Visualization.}
To analyze group movement patterns, we developed a heatmap visualization of the mean speed. For each grid cell $(i, j)$ in a $1024 \times 1024$ discretizations of the space, the average movement velocity was calculated as follows:
$$
\bar{v}_{i,j} = \frac{\sum_{p=1}^{P} \sum_{t=1}^{T} v_{p,t} \cdot \delta_{i,j}(x_{p,t}) \cdot \Delta t}{\sum_{p=1}^{P} \sum_{t=1}^{T} \delta_{i,j}(x_{p,t}) \cdot \Delta t}
$$

where $v_{p,t} = \| x_{p,t+1} - x_{p,t} \| \cdot \text{fps}$ is the velocity of participant $p$ at time $t$, and $\delta_{i,j}$ equals 1 when position $x_{p,t}$ falls within cell $(i, j)$. Gaussian smoothing ($\sigma = 0.01\text{m}$) was applied to create continuous representations. Consistent color scaling across conditions enables direct comparison of spatial usage patterns.

\subsection{Results of User Study 2: Group Sessions}
For each team, we aggregated their movements over the entire task sequence and visualized them as heatmaps. We observed salient differences in the spatial usage patterns. We report the detailed findings below, integrating these observed movement behaviors (Figure~\ref{fig:13}) with insights from participant interviews.

\subsubsection{Functional Visibility and Physical Accessibility}
\begin{figure*}[t]
\centering
\includegraphics[width=0.90\textwidth]{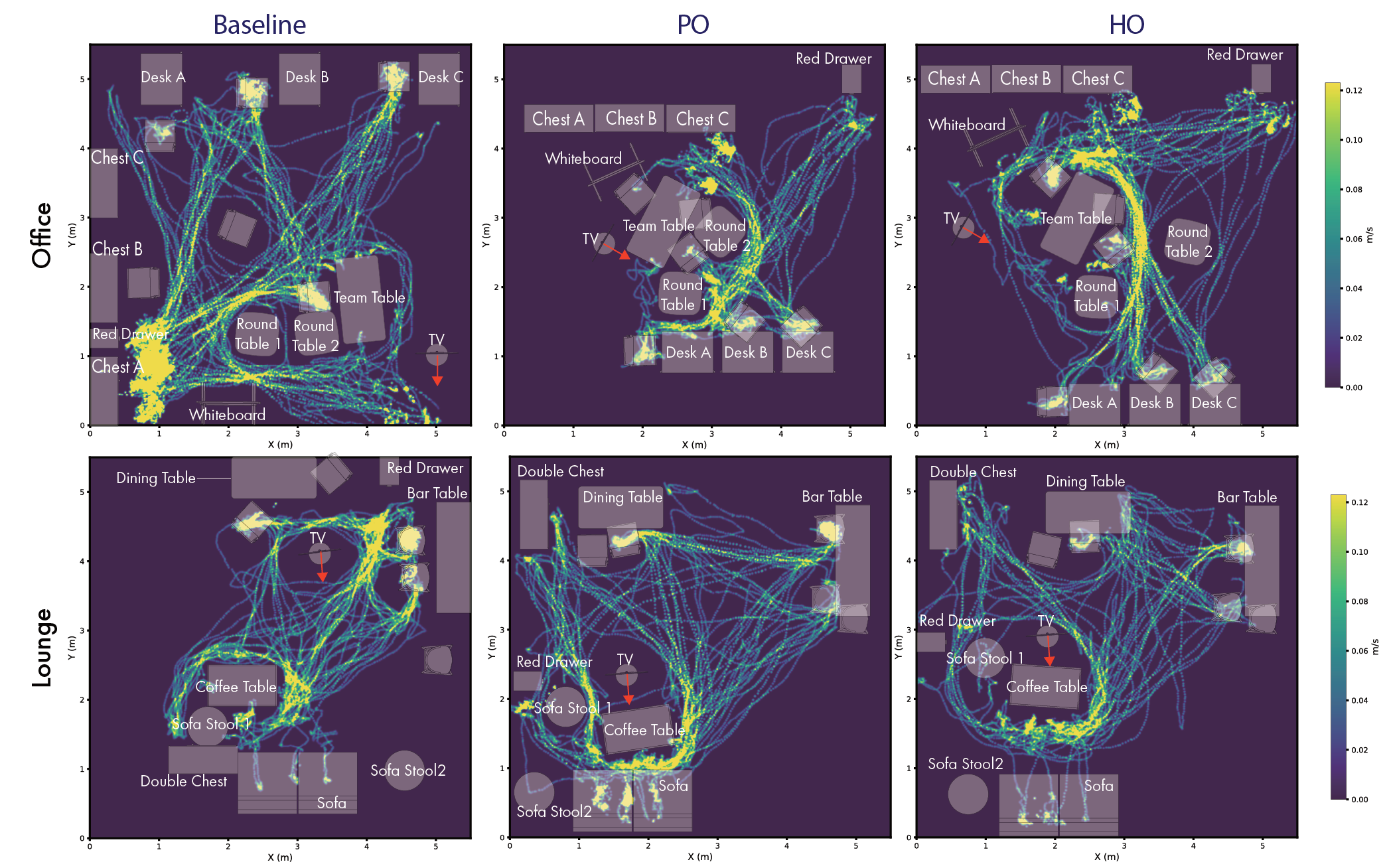}
\caption{\textbf{The Visualization of Mean speed heatmap of Team 1: P21-23 (Office; Top) and Team 6 : P36-38 (Lounge; Bottom)}}
\Description{A diagram visualizes mean speed heatmaps for Team 1 in the office and Team 6 in the lounge, comparing spatial usage patterns across three layout conditions (Baseline, PO, HO). Heatmap intensity indicates movement velocity, with yellow representing high-speed areas and dark purple showing low-speed or stationary zones. In the office, baseline reveals unnecessary movement trajectories generated to access the TV, which is isolated from spatial relationships. PO displays concentrated movement patterns resembling bottlenecks, caused by a lack of secured path clearance that restricts trajectory diversity. In contrast, HO demonstrates fluid movement patterns with efficient circulation around the Team Table and clear access to all workstations. In the lounge, baseline shows a bottleneck between the Red Drawer and Bar Table with fragmented circulation patterns, PO exhibits constrained paths between the Sofa Stool, Coffee Table, and Sofa with limited passage width, and HO enables smooth trajectories with adequate clearance for furniture interaction and unobstructed circulation between functional zones.}
\label{fig:13}
\end{figure*}

In the baseline condition, geometric placement often failed to support functional visibility and access, leading to unnecessary movement. In the office, the TV orientation forced repeated back-and-forth movements to read the screen, whereas in PO and HO layouts, no additional movements were required (Figure~\ref{fig:13}). In interviews, 7 of 9 participants (P23--P29) identified the TV as the most inconvenient asset; P28 noted, ``\textit{I wanted to rotate the TV toward the table where people sit,}'' and P29 remarked, ``\textit{The path required just to see the TV screen was too long.}''

Similar issues emerged in the lounge, where participants identified the TV and obstructed Double Chest as major pain points. Three participants (P30, P31, P36) reported the TV was too distant and only visible from a diagonal angle. Furthermore, access to the Double Chest was obstructed by Sofa Stool 1, which blocked frontal access and forced side-only interaction. Four participants (P30--P32, P38) identified this obstruction as the primary source of inconvenience, explicitly noting that the stool hindered both material retrieval and circulation.

\subsubsection{Trajectory Continuity and Operational Clearance}
Focusing on the distinction between HO and PO, trajectory visualizations revealed that PO layouts often caused trajectory breaks and operational bottlenecks due to cramped configurations (Figure~\ref{fig:13}).

In the office, HO showed continuous movement, whereas PO exhibited breaks between Chest B and the whiteboard, and between Round Table 1 and Desk 1. As P28 reported about PO, ``\textit{Because the Round Table was attached to the Team Table, sitting was uncomfortable, and during standing meetings, I had to walk around excessively.}''

Similar patterns were observed in the lounge. For the Double Chest, HO provided adequate human-object manipulation space, yielding smooth, front-facing trajectories. In PO, the Dining Table encroached on the chest's manipulation space, narrowing the movement corridor; P22 observed, ``\textit{The double chest door didn't open fully, and a bottleneck formed in front of it.}'' Four participants (P30, P36--P38) described the distance between the upper chest and dining table as too narrow, making it difficult to use the chest and sit in nearby chairs. In contrast, participants described HO as more spacious and better at using the available space, emphasizing that moving between the upper chest and other work areas felt unconstrained. P30 specifically noted that opening the upper chest was improved and rated HO as the most favorable layout.

\subsubsection{Summary}
Across group sessions, the HO condition consistently provided better group-level usability than both the baseline and PO layouts, which is consistent with both the trajectory visualizations and participant reports.

\begin{itemize}
    \item \textbf{Functional Visibility and Accessibility:} HO ensured that shared resources such as the TV, sofa, and storage furniture were accessible at appropriate distances and directions. In contrast, baseline layouts frequently compromised visibility or physically blocked access, leading to unnecessary movement.
    \item \textbf{Trajectory Continuity and Operational Clearance:} HO supported continuous movement patterns around key work areas, minimizing the bottlenecks observed in PO. Although PO provided basic passage width, it failed to account for operational encroachment (e.g., furniture blocking storage access), resulting in trajectory breaks and user complaints.
\end{itemize}

These qualitative observations complement our quantitative findings, demonstrating that anthropometrically grounded layouts not only improve individual task performance, but also facilitate more natural and efficient group interactions in shared spaces.

\section{Discussion}

Our findings demonstrate that behavior-aware anthropometric constraints substantially improve layout usability beyond semantic plausibility. We discuss implications across four key areas: functional relationships in generated layouts, operational clearance requirements, applications to XR environments, and limitations with future research directions.

\subsection{Functional Relationships in Behavior-Aware Layouts}
The comparison between the semantic-driven baseline and HO demonstrates that semantic plausibility does not guarantee usability. Although the semantic-driven baseline correctly grouped related objects---such as positioning a coffee table near a sofa, it often failed to provide the physical clearance necessary for interaction.

Our two user studies on human-operational usability clearly exposed these limitations, revealing substantial functional disconnections in the baseline layouts. For example, baselines frequently placed TVs at awkward viewing angles or positioned tables blocking whiteboard access---arrangements that were semantically valid but practically unusable. In contrast, our behavior-aware layouts explicitly inferred the operational envelope required for human actions. HO preserved these functional relationships, allowing participants to maintain natural workflows without repeatedly walking to the TV or viewing it from an awkward diagonal angle. Overall, these results show that semantic-driven approaches alone are insufficient to capture human behavioral patterns or functional relationships between objects and that behavior-aware constraints materially improve the usability of generated spaces.

\subsection{Operational Clearance for Manipulation}
The comparison between PO and HO highlights a critical distinction: passage clearance is not equivalent to operational clearance. Although both conditions use anthropometric data, they differ in the definition of clearance. PO ensures only passage clearance based on body depth and width, whereas HO additionally enforces operational clearance for manipulation and interaction.

This distinction leads to substantial behavioral differences. Compared to PO, HO significantly reduced compensatory movements and trajectory interruptions, eliminating the need for users to reposition furniture or make detours around obstacles. The higher volumetric occupancy under HO quantitatively confirms that securing space for passing through does not automatically guarantee space for operating. Particularly in the group study, the lack of operational constraints in the PO caused practical bottlenecks: objects were placed too close together, obstructing seating at the round table or preventing chest doors from opening fully due to encroaching furniture. These issues appear as trajectory discontinuities in otherwise collision-free layouts. These comparisons confirm that even with the same anthropometric data, there is a difference between designing for mere passage clearance and designing for operational clearance. These findings suggest practical guidance for constraint selection: operational clearance (HO) provides measurable benefits when layouts involve frequent object manipulation—such as opening drawers, adjusting furniture, or accessing storage—whereas passage-only constraints (PO) may suffice for circulation-focused spaces with minimal interaction requirements. By clarifying when anthropometric-aware operational constraints yield usability gains, our work helps practitioners decide where to invest additional design effort.

\subsection{Applications to 3D Environments}
Our framework encodes anthropometric considerations as differentiable constraints, which opens opportunities for applications beyond static physical layout generation. We discuss three potential directions where behavior-aware scene generation could provide value, while noting that empirical validation in these domains remains for future work.

\paragraph{Adaptation to Device-Augmented Interactions}
In XR environments, users often interact through controllers or perform wide gestural commands, effectively extending their kinematic volume beyond their physical body dimensions~\cite{ahuja2022controllerpose, 
scavarelli2017vr}. Our framework could be extended to such device-specific operational envelopes, potentially 
enabling layouts that account for the additional clearance required during controller-based or gestural interactions. This represents an extension of our anthropometric parameterization, though the specific requirements would need to be validated through user studies in XR settings.

\paragraph{Enhancing Embodied Interaction Fidelity in XR}
Maintaining immersion in XR depends on minimizing the gap between user intent and virtual constraints~\cite{tao2023embodying}. Notably, virtual environments allow instant spatial modifications without physical effort, making real-time anthropometric adaptation feasible---a capability impractical in physical settings. Applying our framework to XR implies that generated environments could help prevent the interaction mismatches---such as users having to overextend to reach a control or seeing their virtual hands penetrate a table surface. By optimizing layouts based on the specific anthropometric profiles of users (or avatars), our framework has the potential to support natural, collision-free workflows in virtual spaces, thereby contributing to higher embodied interaction fidelity without requiring manual post-processing.

\paragraph{Scalable Content Creation for AI agents and Digital Twins}
Finally, for scenarios requiring mass production of synthetic scenes---such as AI agent training, robotic navigation, or digital twin simulations---our method may offer a pathway toward improved behavioral realism. Instead of static asset placement, users can specify high-level activities (e.g., \textit{sitting and retrieving objects}). The system then generates layouts that provide the kinematic space required for these actions, potentially reducing the awkward motions or unnatural inverse kinematics often observed in simulations where virtual agents or robots struggle to navigate functionally disconnected spaces.

\subsection{Limitations and Future Works}
Our study focuses on investigating whether anthropometric and behavioral data can improve usability in computational scene generation.  With this scope in mind, we discuss several limitations and future directions below.

First, regarding generalizability and validation scope, our technical validation covered 20 scenes and user studies focused on 6 scenes. Additionally, while our participant pool provided sufficient power to detect usability improvements within these contexts, the sample size limits the ability to make broad claims about diverse populations. Future research should conduct large-scale evaluations and expand validation to complex, highly regulated environments such as kitchens, bathrooms, or retail spaces to establish normative design guidelines. Second, our current framework focuses on horizontal arrangements due to the lack of parametric variability in our asset library. Consequently, vertical interactions---such as shelf heights or standing-to-sitting transitions---were not optimized. Future work should incorporate parametric object generation to enable full 3D optimization, allowing the system to adjust furniture heights based on individual anthropometric profiles. Third, regarding multi-user optimization, our current approach prioritized accessibility by using the maximum body dimensions within a group to ensure clearance for all. However, this revealed an inherent trade-off: ensuring geometric clearance for the largest user often resulted in uncomfortable interaction distances for shorter participants (P22). More broadly, this finding illustrates that the appropriate constraint priority may vary depending on the object relationship and intended interaction—some relationships benefit from closer placement for reach, while others require wider spacing for manipulation. Since our primary focus was to examine how behavior and anthropometric information affects layout quality, we did not address adaptively balancing such competing constraints. Future research should explore adaptive constraint prioritization that considers the specific interaction context of each object pair, aligning with universal design principles for diverse user populations. Fourth, our VLM-based constraint inference was validated using representative furniture assets from preprocessed datasets. The system reliably inferred functional descriptions and interaction patterns for the tested furniture assets. However, since the VLM relies on visual cues and common sense, affordances that are not visually apparent (e.g., hidden mechanisms or non-standard opening directions) cannot be inferred without supplementary text information. Generalization to unconventional furniture designs—such as diverse forms, or ambiguous multi-functional pieces—remains unexamined. Finally, our current framework was instantiated and evaluated in physical environments; however, the same anthropometric and behavior-aware constraints can be extended to XR settings. As XR technologies become more prevalent in workplaces and residential environments \cite{kim2024spatialaffordanceawareinteractablesubspace, luo2022should, luo2025documents}, spatial layouts will increasingly need to accommodate physical and virtual interaction zones.

\section{Conclusion}
We presented Behavior-Aware Anthropometric Scene Generation, an approach that augments language-based layout generation with behavioral reasoning and anthropometric grounding, shifting the focus from visual and semantic plausibility to human-operational usability by examining how behavioral and anthropometric information affects layout quality. Our contributions are threefold: First, we introduce a two-stage scene generation framework that translates behavioral reasoning into differentiable spatial constraints, enabling gradient-based optimization of layout usability. Second, we defined explicit spatial constraints for operational clearance and interaction zones, ensuring that layouts accommodate dynamic human actions---such as reaching or opening drawers---rather than merely avoiding static collisions. Finally, we demonstrated through technical validation and user studies that behavior-centric metrics capture how people actually use space beyond conventional collision and boundary scores. Our behavior-aware anthropometric perspective represents a critical step toward functional operational environments in XR, embodied AI, and digital twin systems, where human movement, reach, and interaction become first-class design constraints.

\begin{acks}
This work was supported by the Technology Innovation Program (RS-2025-02317326, Development of AI-Driven Design Generation Technology Based on Designer Intent) funded by the Ministry of Trade, Industry \& Energy (MOTIE, Korea).
\end{acks}

\bibliographystyle{ACM-Reference-Format}
\bibliography{reference}

@String{Computing = "Computing" }

@String{Computer = "{IEEE} Computer" }

@String{Springer = "Springer-Verlag" }

@inproceedings{gu2018ava,
  title={Ava: A video dataset of spatio-temporally localized atomic visual actions},
  author={Gu, Chunhui and Sun, Chen and Ross, David A and Vondrick, Carl and Pantofaru, Caroline and Li, Yeqing and Vijayanarasimhan, Sudheendra and Toderici, George and Ricco, Susanna and Sukthankar, Rahul and others},
  booktitle={Proceedings of the IEEE conference on computer vision and pattern recognition},
  pages={6047--6056},
  year={2018},
  doi={https://doi.org/10.1109/cvpr.2018.00633}
}

@inproceedings{bojanic2024pose,
  title={Pose-independent 3d anthropometry from sparse data},
  author={Bojani{\'c}, David and Wuhrer, Stefanie and Petkovi{\'c}, Tomislav and Pribani{\'c}, Tomislav},
  booktitle={European Conference on Computer Vision},
  pages={237--256},
  year={2024},
  organization={Springer},
  doi={https://doi.org/10.1007/978-3-031-91575-8_15}
}

@article{zhai2023commonscenes,
  title={Commonscenes: Generating commonsense 3d indoor scenes with scene graph diffusion},
  author={Zhai, Guangyao and {\"O}rnek, Evin P{\i}nar and Wu, Shun-Cheng and Di, Yan and Tombari, Federico and Navab, Nassir and Busam, Benjamin},
  journal={Advances in Neural Information Processing Systems},
  volume={36},
  pages={30026--30038},
  year={2023},
  doi={https://doi.org/10.48550/arXiv.2305.16283}
}

@inproceedings{deitke2023objaverse,
  title={Objaverse: A universe of annotated 3d objects},
  author={Deitke, Matt and Schwenk, Dustin and Salvador, Jordi and Weihs, Luca and Michel, Oscar and VanderBilt, Eli and Schmidt, Ludwig and Ehsani, Kiana and Kembhavi, Aniruddha and Farhadi, Ali},
  booktitle={Proceedings of the IEEE/CVF conference on computer vision and pattern recognition},
  pages={13142--13153},
  year={2023},
  doi={https://doi.org/10.1109/cvpr52729.2023.01263}
}

@inproceedings{luo2023pearl,
title={Pearl: Physical environment based augmented reality lenses for in-situ human movement analysis},
author={Luo, Weizhou and Yu, Zhongyuan and Rzayev, Rufat and Satkowski, Marc and Gumhold, Stefan and McGinity, Matthew and Dachselt, Raimund},
booktitle={Proceedings of the 2023 CHI Conference on Human Factors in Computing Systems},
pages={1--15},
year={2023},
doi={https://doi.org/10.1145/3544548.3580715}
}

@inproceedings{nguyen2024adaptive,
  title={The adaptive architectural layout: How the control of a semi-autonomous mobile robotic partition was shared to mediate the environmental demands and resources of an open-plan office},
  author={Nguyen, Binh Vinh Duc and Vande Moere, Andrew},
  booktitle={Proceedings of the 2024 CHI Conference on Human Factors in Computing Systems},
  pages={1--20},
  year={2024},
  doi={https://doi.org/10.1145/3613904.3642465}
}

@article{julia2021waiting,
  title={Waiting room physical environment and outpatient experience: The spatial user experience model as analytical tool},
  author={Juli{\'a} Nehme, Bego{\~n}a and Torres Irribarra, David and Cumsille, Patricio and Yoon, So--Yeon},
  journal={Journal of Interior Design},
  volume={46},
  number={4},
  pages={27--48},
  year={2021},
  publisher={SAGE Publications Sage CA: Los Angeles, CA},
  doi={https://doi.org/10.1111/joid.12205}
}

@article{wang2022humanise,
title={Humanise: Language-conditioned human motion generation in 3d scenes},
author={Wang, Zan and Chen, Yixin and Liu, Tengyu and Zhu, Yixin and Liang, Wei and Huang, Siyuan},
journal={Advances in Neural Information Processing Systems},
volume={35},
pages={14959--14971},
year={2022},
doi={https://doi.org/10.48550/arXiv.2210.09729}
}

@article{julia2020spatial,
  title={Spatial user experience: A multidisciplinary approach to assessing physical settings},
  author={Juli{\'a} Nehme, Bego{\~n}a and Rodr{\'\i}guez, Eugenio and Yoon, So--Yeon},
  journal={Journal of Interior Design},
  volume={45},
  number={3},
  pages={7--25},
  year={2020},
  publisher={SAGE Publications Sage CA: Los Angeles, CA},
  doi={https://doi.org/10.1111/joid.12177}
}

@inproceedings{yang2024physcene,
title={Physcene: Physically interactable 3d scene synthesis for embodied ai},
author={Yang, Yandan and Jia, Baoxiong and Zhi, Peiyuan and Huang, Siyuan},
booktitle={Proceedings of the IEEE/CVF Conference on Computer Vision and Pattern Recognition},
pages={16262--16272},
year={2024},
doi={https://doi.org/10.1109/cvpr52733.2024.01539}
}

@article{sun2024haisor,
title={Haisor: Human-aware indoor scene optimization via deep reinforcement learning},
author={Sun, Jia-Mu and Yang, Jie and Mo, Kaichun and Lai, Yu-Kun and Guibas, Leonidas and Gao, Lin},
journal={ACM Transactions on Graphics},
volume={43},
number={2},
pages={1--17},
year={2024},
publisher={ACM New York, NY, USA},
doi={https://doi.org/10.1145/3632947}
}

@inproceedings{luo2025documents,
title={Documents in Your Hands: Exploring Interaction Techniques for Spatial Arrangement of Augmented Reality Documents},
author={Luo, Weizhou and Ellenberg, Mats Ole and Satkowski, Marc and Dachselt, Raimund},
booktitle={Proceedings of the 2025 CHI Conference on Human Factors in Computing Systems},
pages={1--22},
year={2025},
doi={https://doi.org/10.1145/3706598.3713518}
}

@InProceedings{Tang_2024_CVPR,
author = {Tang, Jiapeng and Nie, Yinyu and Markhasin, Lev and Dai, Angela and Thies, Justus and Nie{\ss}ner, Matthias},
title = {DiffuScene: Denoising Diffusion Models for Generative Indoor Scene Synthesis},
booktitle = {Proceedings of the IEEE/CVF Conference on Computer Vision and Pattern Recognition (CVPR)},
month = {June},
year = {2024},
pages = {20507-20518},
doi={https://doi.org/10.1109/cvpr52733.2024.01938}
}

@article{dianat2018review,
title={A review of the methodology and applications of anthropometry in ergonomics and product design},
author={Dianat, Iman and Molenbroek, Johan and Castellucci, H{\'e}ctor Ignacio},
journal={Ergonomics},
volume={61},
number={12},
pages={1696--1720},
year={2018},
publisher={Taylor \& Francis},
doi={https://doi.org/10.1080/00140139.2018.1502817 }
}

@inproceedings{luo2022should,
title={Where should we put it? layout and placement strategies of documents in augmented reality for collaborative sensemaking},
author={Luo, Weizhou and Lehmann, Anke and Widengren, Hjalmar and Dachselt, Raimund},
booktitle={Proceedings of the 2022 CHI Conference on Human Factors in Computing Systems},
pages={1--16},
year={2022},
doi={https://doi.org/10.1145/3491102.3501946}
}

@inproceedings{wang2021sceneformer,
title={Sceneformer: Indoor scene generation with transformers},
author={Wang, Xinpeng and Yeshwanth, Chandan and Nie{\ss}ner, Matthias},
booktitle={2021 International Conference on 3D Vision (3DV)},
pages={106--115},
year={2021},
organization={IEEE},
doi={https://doi.org/10.1109/3dv53792.2021.00021}
}

@article{paschalidou2021atiss,
title={Atiss: Autoregressive transformers for indoor scene synthesis},
author={Paschalidou, Despoina and Kar, Amlan and Shugrina, Maria and Kreis, Karsten and Geiger, Andreas and Fidler, Sanja},
journal={Advances in Neural Information Processing Systems},
volume={34},
pages={12013--12026},
year={2021},
doi={https://doi.org/10.48550/arXiv.2110.03675}
}

@inproceedings{sun2025forest2seq,
title={Forest2seq: Revitalizing order prior for sequential indoor scene synthesis},
author={Sun, Qi and Zhou, Hang and Zhou, Wengang and Li, Li and Li, Houqiang},
booktitle={European Conference on Computer Vision},
pages={251--268},
year={2025},
organization={Springer},
doi={https://doi.org/10.1007/978-3-031-72698-9_15}
}

@article{liu2023openshape,
  title={Openshape: Scaling up 3d shape representation towards open-world understanding},
  author={Liu, Minghua and Shi, Ruoxi and Kuang, Kaiming and Zhu, Yinhao and Li, Xuanlin and Han, Shizhong and Cai, Hong and Porikli, Fatih and Su, Hao},
  journal={Advances in neural information processing systems},
  volume={36},
  pages={44860--44879},
  year={2023},
  doi={https://doi.org/10.48550/arXiv.2305.10764}
}

@inproceedings{sun2025layoutvlm,
  title={Layoutvlm: Differentiable optimization of 3d layout via vision-language models},
  author={Sun, Fan-Yun and Liu, Weiyu and Gu, Siyi and Lim, Dylan and Bhat, Goutam and Tombari, Federico and Li, Manling and Haber, Nick and Wu, Jiajun},
  booktitle={Proceedings of the Computer Vision and Pattern Recognition Conference},
  pages={29469--29478},
  year={2025},
  doi={https://doi.org/10.1109/cvpr52734.2025.02744}
}

@inproceedings{achlioptas2020referit3d,
  title={Referit3d: Neural listeners for fine-grained 3d object identification in real-world scenes},
  author={Achlioptas, Panos and Abdelreheem, Ahmed and Xia, Fei and Elhoseiny, Mohamed and Guibas, Leonidas},
  booktitle={European conference on computer vision},
  pages={422--440},
  year={2020},
  organization={Springer},
  doi={https://doi.org/10.1007/978-3-030-58452-8_25}
}

@inproceedings{ramakers2023measurement,
title={Measurement patterns: User-oriented strategies for dealing with measurements and dimensions in making processes},
author={Ramakers, Raf and Leen, Danny and Kim, Jeeeun and Luyten, Kris and Houben, Steven and Veuskens, Tom},
booktitle={Proceedings of the 2023 CHI Conference on Human Factors in Computing Systems},
pages={1--17},
year={2023},
doi={https://doi.org/10.1145/3544548.3581157}
}

@article{lin2024instructscene,
  title={Instructscene: Instruction-driven 3d indoor scene synthesis with semantic graph prior},
  author={Lin, Chenguo and Mu, Yadong},
  journal={arXiv preprint arXiv:2402.04717},
  year={2024},
  doi={https://doi.org/10.48550/arXiv.2402.04717}
}

@inproceedings{shuai2022multinb,
title={Novel View Synthesis of Human Interactions from Sparse
Multi-view Videos},
author={Shuai, Qing and Geng, Chen and Fang, Qi and Peng, Sida and Shen, Wenhao and Zhou, Xiaowei and Bao, Hujun},
booktitle={SIGGRAPH Conference Proceedings},
year={2022},
doi={https://doi.org/10.1145/3528233.3530704}
}

@inproceedings{lee2016posing,
title={Posing and acting as input for personalizing furniture},
author={Lee, Bokyung and Cho, Minjoo and Min, Joonhee and Saakes, Daniel},
booktitle={Proceedings of the 9th Nordic Conference on Human-Computer Interaction},
pages={1--10},
year={2016},
doi={https://doi.org/10.1145/2971485.2971487}
}

@article{chen2025exploring,
title={Exploring the role of Mixed Reality on Design Representations to Enhance User-Involved Co-Design Communication},
author={Chen, Pei and Wang, Kexing and Liu, Lianyan and Liu, Xuanhui and Zhang, Hongbo and Teng, Zhuyu and Sun, Lingyun},
journal={Proceedings of the ACM on Human-Computer Interaction},
volume={9},
number={2},
pages={1--29},
year={2025},
publisher={ACM New York, NY, USA},
doi={https://doi.org/10.1145/3710979}
}

@article{SMPL:2015,
  title = {{SMPL}: A Skinned Multi-Person Linear Model},
  author = {Loper, Matthew and Mahmood, Naureen and Romero, Javier and Pons-Moll, Gerard and Black, Michael J.},
  journal = {ACM Trans. Graphics (Proc. SIGGRAPH Asia)},
  year = {2015},
  doi={https://doi.org/10.1145/3596711.3596800}
}

@inproceedings{ccelen2024design,
  title={I-design: Personalized llm interior designer},
  author={{\c{C}}elen, Ata and Han, Guo and Schindler, Konrad and Van Gool, Luc and Armeni, Iro and Obukhov, Anton and Wang, Xi},
  booktitle={European Conference on Computer Vision},
  pages={217--234},
  year={2024},
  organization={Springer},
  doi={https://doi.org/10.1007/978-3-031-92387-6_17}
}

@inproceedings{wang2024move,
title={Move as You Say Interact as You Can: Language-guided Human Motion Generation with Scene Affordance},
author={Wang, Zan and Chen, Yixin and Jia, Baoxiong and Li, Puhao and Zhang, Jinlu and Zhang, Jingze and Liu, Tengyu and Zhu, Yixin and Liang, Wei and Huang, Siyuan},
booktitle={Proceedings of the IEEE/CVF Conference on Computer Vision and Pattern Recognition},
pages={433--444},
year={2024},
doi={https://doi.org/10.1109/cvpr52733.2024.00049}
}

@book{panero1979human,
  title={Human Dimension and Interior Space: A Source Book of Design Reference Standards},
  author={Panero, Julius and Zelnik, Martin},
  year={1979},
  publisher={Watson-Guptill},
  address={New York}
}

@book{damon1966human,
title={The human body in equipment design},
author={Damon, Albert and Stoudt, Howard W and McFarland, Ross A},
year={1966},
publisher={Harvard University Press},
doi={https://doi.org/10.4159/harvard.9780674491892}
}

@book{ching2023architecture,
  title={Architecture: Form, space, and order},
  author={Ching, Francis DK},
  year={2023},
  publisher={John Wiley \& Sons}
}

@article{viviani2018accuracy,
  title={Accuracy, precision and reliability in anthropometric surveys for ergonomics purposes in adult working populations: A literature review},
  author={Viviani, Carlos and Arezes, PM and Braganca, Sara and Molenbroek, Johan and Dianat, Iman and Castellucci, HI},
  journal={International Journal of Industrial Ergonomics},
  volume={65},
  pages={1--16},
  year={2018},
  publisher={Elsevier},
doi = {https://doi.org/10.1016/j.ergon.2018.01.012}
}

@article{qu2024gpt,
title={GPT-Connect: Interaction between Text-Driven Human Motion Generator and 3D Scenes in a Training-free Manner},
author={Qu, Haoxuan and Guo, Ziyan and Liu, Jun},
journal={arXiv preprint arXiv:2403.14947},
year={2024},
doi={https://doi.org/10.48550/arXiv.2403.14947}
}

@inproceedings{ding2023task,
title={Task and motion planning with large language models for object rearrangement},
author={Ding, Yan and Zhang, Xiaohan and Paxton, Chris and Zhang, Shiqi},
booktitle={2023 IEEE/RSJ International Conference on Intelligent Robots and Systems (IROS)},
pages={2086--2092},
year={2023},
organization={IEEE},
doi={https://doi.org/10.1109/iros55552.2023.10342169}
}

@inproceedings{wu2025human,
  title={Human-object interaction from human-level instructions},
  author={Wu, Zhen and Li, Jiaman and Xu, Pei and Liu, C Karen},
  booktitle={Proceedings of the IEEE/CVF International Conference on Computer Vision},
  pages={11176--11186},
  year={2025},
doi={https://doi.org/10.48550/arXiv.2406.17840}
}

@article{su2023scene,
title={Scene-aware Activity Program Generation with Language Guidance Supplementary Material},
author={Su, Zejia and Fan, Qingnan and Chen, Xuelin and Van Kaick, Oliver and Huang, Hui and Hu, Ruizhen},
journal={ACM Trans. Graph},
volume={42},
number={6},
year={2023},
doi={https://doi.org/10.1145/3618338}
}

@book{van1972human,
title={Human engineering guide to equipment design},
author={Van Cott, Harold P and Kinkade, Robert G},
year={1972},
publisher={Department of Defense}
}

@book{nasa1978anthropometric,
title={Anthropometric Source Book. Volume 2: A Handbook of Anthropometric Data},
author={NASA.},
year={1978},
publisher={Webb Associates yellow springs oh}
}

@misc{2020mmaction2,
  title={OpenMMLab's Next Generation Video Understanding Toolbox and Benchmark},
  author={MMAction2 Contributors},
  howpublished = {\url{https://github.com/open-mmlab/mmaction2}},
  year={2020}
}

@inproceedings{lee2018interactive,
  title={Interactive and situated guidelines to help users design a personal desk that fits their bodies},
  author={Lee, Bokyung and Shin, Joongi and Bae, Hyoshin and Saakes, Daniel},
  booktitle={Proceedings of the 2018 designing interactive systems conference},
  pages={637--650},
  year={2018},
  doi={https://doi.org/10.1145/3196709.3196725}
}

@misc{hu2024mixeddiffusion3dindoor,
      title={Mixed Diffusion for 3D Indoor Scene Synthesis}, 
      author={Siyi Hu and Diego Martin Arroyo and Stephanie Debats and Fabian Manhardt and Luca Carlone and Federico Tombari},
      year={2024},
      eprint={2405.21066},
      archivePrefix={arXiv},
      primaryClass={cs.CV},
      url={https://arxiv.org/abs/2405.21066}, 
doi={https://doi.org/10.48550/arXiv.2405.21066}
}

@inproceedings{
    feng2023layoutgpt,
    title={Layout{GPT}: Compositional Visual Planning and Generation with Large Language Models},
    author={Weixi Feng and Wanrong Zhu and Tsu-Jui Fu and Varun Jampani and Arjun Reddy Akula and Xuehai He and S Basu and Xin Eric Wang and William Yang Wang},
    booktitle={Thirty-seventh Conference on Neural Information Processing Systems},
    year={2023},
    url={https://openreview.net/forum?id=Xu8aG5Q8M3}
}

@misc{yang2024holodecklanguageguidedgeneration,
      title={Holodeck: Language Guided Generation of 3D Embodied AI Environments}, 
      author={Yue Yang and Fan-Yun Sun and Luca Weihs and Eli VanderBilt and Alvaro Herrasti and Winson Han and Jiajun Wu and Nick Haber and Ranjay Krishna and Lingjie Liu and Chris Callison-Burch and Mark Yatskar and Aniruddha Kembhavi and Christopher Clark},
      year={2024},
      eprint={2312.09067},
      archivePrefix={arXiv},
      primaryClass={cs.CV},
      url={https://arxiv.org/abs/2312.09067},
doi={https://doi.org/10.1109/cvpr52733.2024.01536}
}

@misc{kim2024spatialaffordanceawareinteractablesubspace,
      title={Spatial Affordance-aware Interactable Subspace Allocation for Mixed Reality Telepresence}, 
      author={Dooyoung Kim and Seonji Kim and Selin Choi and Woontack Woo},
      year={2024},
      eprint={2408.04297},
      archivePrefix={arXiv},
      primaryClass={cs.ET},
      url={https://arxiv.org/abs/2408.04297}, 
      doi={https://doi.org/10.1109/ismar62088.2024.00142}
}

@inproceedings{scavarelli2017vr,
  doi={https://doi.org/10.1145/3027063.3053180},
  title={Vr collide! comparing collision-avoidance methods between co-located virtual reality users},
  author={Scavarelli, Anthony and Teather, Robert J},
  booktitle={Proceedings of the 2017 CHI conference extended abstracts on human factors in computing systems},
  pages={2915--2921},
  year={2017}
}

@inproceedings{tao2023embodying,
  doi={https://doi.org/10.1145/3544548.3580979},
  title={Embodying physics-aware avatars in virtual reality},
  author={Tao, Yujie and Wang, Cheng Yao and Wilson, Andrew D and Ofek, Eyal and Gonzalez-Franco, Mar},
  booktitle={Proceedings of the 2023 CHI Conference on Human Factors in Computing Systems},
  pages={1--15},
  year={2023}
}

@inproceedings{ahuja2022controllerpose,
  doi={https://doi.org/10.1145/3491102.3502105 },
  title={Controllerpose: inside-out body capture with VR controller cameras},
  author={Ahuja, Karan and Shen, Vivian and Fang, Cathy Mengying and Riopelle, Nathan and Kong, Andy and Harrison, Chris},
  booktitle={Proceedings of the 2022 CHI Conference on Human Factors in Computing Systems},
  pages={1--13},
  year={2022}
}

\end{document}